\documentclass[twocolumn]{aastex631}

\shorttitle{Multiphase CGM}
\shortauthors{Khonde et al.}

\usepackage{newtxtext,newtxmath}

\usepackage[T1]{fontenc}

\DeclareRobustCommand{\VAN}[3]{#2}
\let\VANthebibliography\thebibliography
\def\thebibliography{\DeclareRobustCommand{\VAN}[3]{##3}\VANthebibliography}

\usepackage{graphicx}	
\usepackage[caption=false]{subfig}
\usepackage{amsmath}	
\usepackage{float}
\usepackage{verbatim} 
\usepackage{graphics}
\usepackage{xcolor}

\def\CIVdblt{{\rm C~}\kern 0.1em{\sc iv}~$\lambda\lambda 1548, 1550$}
\def\NVdblt{{\rm N~}\kern 0.1em{\sc v}~$\lambda\lambda 1238, 1242$}
\def\OVIdblt{{\rm O~}\kern 0.1em{\sc vi}~$\lambda\lambda 1031, 1037$}
\def\SVIdblt{{\rm S~}\kern 0.1em{\sc vi}~$\lambda\lambda 933, 944$}
\def\SiIVdblt{{\rm Si~}\kern 0.1em{\sc iv}~$\lambda\lambda 1394, 1403$}
\def\MgIIdblt{{\rm Mg~}\kern 0.1em{\sc ii}~$\lambda\lambda 2796, 2803$ }
\def\AlIIIdblt{{\rm Al~}\kern 0.1em{\sc iii}~$\lambda\lambda 1854, 1862$}
\def\NeVIIIdblt{{\rm Ne~}\kern 0.1em{\sc viii}~$\lambda\lambda770, 780$}
\def\NeV{\hbox{{\rm Ne~}\kern 0.1em{\sc v}}}
\def\NeVI{\hbox{{\rm Ne~}\kern 0.1em{\sc vi}}}
\def\NeVIII{\hbox{{\rm Ne~}\kern 0.1em{\sc viii}}}
\def\OII{\hbox{{\rm O~}\kern 0.1em{\sc ii}}}
\def\OIII{\hbox{{\rm O~}\kern 0.1em{\sc iii}}}
\def\OIV{\hbox{{\rm O~}\kern 0.1em{\sc iv}}}
\def\OV{\hbox{{\rm O~}\kern 0.1em{\sc v}}}
\def\OVI{\hbox{{\rm O~}\kern 0.1em{\sc vi}}}
\def\OVII{\hbox{{\rm O~}\kern 0.1em{\sc vii}}}
\def\OVIII{\hbox{{\rm O~}\kern 0.1em{\sc viii}}}
\def\NI{\hbox{{\rm N~}\kern 0.1em{\sc i}}}
\def\NII{\hbox{{\rm N~}\kern 0.1em{\sc ii}}}
\def\NIII{\hbox{{\rm N~}\kern 0.1em{\sc iii}}}
\def\NIV{\hbox{{\rm N~}\kern 0.1em{\sc iv}}}
\def\NV{\hbox{{\rm N~}\kern 0.1em{\sc v}}}
\def\NVII{\hbox{{\rm N~}\kern 0.1em{\sc vii}}}
\def\CII{\hbox{{\rm C~}\kern 0.1em{\sc ii}}}
\def\CIII{\hbox{{\rm C~}\kern 0.1em{\sc iii}}}
\def\SiII{\hbox{{\rm Si~}\kern 0.1em{\sc ii}}}
\def\SiIII{\hbox{{\rm Si~}\kern 0.1em{\sc iii}}}
\def\SiIV{\hbox{{\rm Si~}\kern 0.1em{\sc iv}}}
\def\SIV{\hbox{{\rm S~}\kern 0.1em{\sc iv}}}
\def\SV{\hbox{{\rm S~}\kern 0.1em{\sc v}}}
\def\SVI{\hbox{{\rm S~}\kern 0.1em{\sc vi}}}
\def\SiI{\hbox{{\rm Si~}\kern 0.1em{\sc i}}}
\def\PII{\hbox{{\rm P~}\kern 0.1em{\sc ii}}}
\def\AlI{\hbox{{\rm Al~}\kern 0.1em{\sc i}}}
\def\AlII{\hbox{{\rm Al~}\kern 0.1em{\sc ii}}}
\def\AlIII{\hbox{{\rm Al~}\kern 0.1em{\sc iii}}}
\def\CaI{\hbox{{\rm Ca~}\kern 0.1em{\sc i}}}
\def\CaII{\hbox{{\rm Ca~}\kern 0.1em{\sc ii}}}
\def\CrII{\hbox{{\rm Cr~}\kern 0.1em{\sc ii}}}
\def\CII{\hbox{{\rm C~}\kern 0.1em{\sc ii}}}
\def\CIII{\hbox{{\rm C~}\kern 0.1em{\sc iii}}}
\def\CIV{\hbox{{\rm C~}\kern 0.1em{\sc iv}}}
\def\CV{\hbox{{\rm C}\kern 0.1em{\sc v}}}
\def\MgX{\hbox{{\rm Mg}\kern 0.1em{\sc x}}}
\def\MgI{\hbox{{\rm Mg}\kern 0.1em{\sc i}}}
\def\MgII{\hbox{{\rm Mg}\kern 0.1em{\sc ii}}}
\def\FeII{\hbox{{\rm Fe~}\kern 0.1em{\sc ii}}}
\def\FeIII{\hbox{{\rm Fe~}\kern 0.1em{\sc iii}}}
\def\TiII{\hbox{{\rm Ti~}\kern 0.1em{\sc ii}}}
\def\NaI{\hbox{{\rm Na~}\kern 0.1em{\sc i}}}
\def\SII{\hbox{{\rm S}\kern 0.1em{\sc ii}}}
\def\H{\hbox{{\rm H~}}}
\def\HI{\hbox{{\rm H}{\sc \,i}}}
\def\HeI{\hbox{{\rm He~}\kern 0.1em{\sc i}}}
\def\HII{\hbox{{\rm H~}\kern 0.1em{\sc ii}}}
\def\Lya{\hbox{{\rm Ly}\kern 0.1em$\alpha$}}
\def\Lyb{\hbox{{\rm Ly}\kern 0.1em$\beta$}}
\def\Lyg{\hbox{{\rm Ly}\kern 0.1em$\gamma$}}
\def\Lyth{\hbox{{\rm Ly}\kern 0.1em$\theta$}}
\def\Lyfive{\hbox{{\rm Ly}\kern 0.1em$5$}}
\def\Lysix{\hbox{{\rm Ly}\kern 0.1em$6$}}
\def\Lyseven{\hbox{{\rm Ly}\kern 0.1em$7$}}
\def\Lyeight{\hbox{{\rm Ly}\kern 0.1em$8$}}
\def\Lynine{\hbox{{\rm Ly}\kern 0.1em$9$}}
\def\Lyten{\hbox{{\rm Ly}\kern 0.1em$10$}}
\def\MnII{\hbox{{\rm Mn~}\kern 0.1em{\sc ii}}}
\def\kms{\hbox{km~s$^{-1}$}}
\def\cmsq{\hbox{cm$^{-2}$}}
\def\cc{\hbox{cm$^{-3}$}}

\newcommand{\CLOUDY}{\textsc{cloudy}}

\begin{document}

\title{Lyman Limit System with {\OVI} in the Circumgalactic Environment of a Pair of Galaxies}

\author[0009-0006-6307-7605]{Dheerajkumar Khonde}
\affiliation{Indian Institute of Space Science and Technology, Thiruvananthapuram 695 547, Kerala, India}
\affiliation{Physical Research Laboratory, Ahmedabad 380 009, Gujarat , India}
\

\author{Purvi Udhwani}
\affiliation{Indian Institute of Space Science and Technology, Thiruvananthapuram 695 547, Kerala, India}
\affiliation{Aryabhatta Research Institute of Observational Sciences, Manora Peak, Nainital 263 129, Uttarakhand, India}

\author{Anand Narayanan}
\affiliation{Indian Institute of Space Science and Technology, Thiruvananthapuram 695 547, Kerala, India}

\author{Sowgat Muzahid}
\affiliation{The Inter-University Centre for Astronomy and Astrophysics, Ganeshkhind, Pune 411 007, Maharashtra, India }

\author{Vikram Khaire}
\affiliation{Indian Institute of Space Science and Technology, Thiruvananthapuram 695 547, Kerala, India}
\affiliation{Physics Department, Broida Hall, University of California Santa Barbara, Santa Barbara, CA 93106-9530, USA}

\author{Martin Wendt}
\affiliation{Institut für Physik und Astronomie, Universität Potsdam,Karl-Liebknecht-Str. 24/25, 14476 Golm, Germany}

\begin{abstract}

We report on the analysis of a multiphase Lyman limit system (LLS) at $z=0.39047$ identified towards the background quasar FBQS~J$0209-0438$. The {\OVI} doublet lines associated with this absorber have a different profile from the low ionization metals and the {\HI}. The {\Lya} has a very broad {\HI} ($b \approx 150$ {\kms}) component well-aligned with one of the {\OVI} components. The Doppler $b$-parameters for the broad {\HI} and {\OVI} indicate gas with $T= (0.8-2.0)\times 10^6$~K, and a total hydrogen column density that is an order of magnitude larger than the cooler phase of gas responsible for the LLS. Observations by $VLT$/MUSE show two moderately star-forming galaxies within $\rho \lesssim 105$~kpc, and $|\Delta v|\lesssim 130$ {\kms} of the absorber, one of them a dwarf galaxy ($M_*\approx 10^6$~M$_\odot$) overlapping with the quasar PSF, and the other a larger galaxy ($R_{1/2}\approx 4$~kpc) with $M_*\approx 3\times 10^{10}$~M$_\odot$ and $M_h\approx 7\times 10^{11}$~M$_\odot$, and the dwarf galaxy within its virial radius. Though the absorption is aligned with the extended major axis of the larger galaxy, the line-of-sight velocity of the absorbing gas is inconsistent with corotating accretion. The metallicity inferred for the LLS is lower than the gas phase [O/H] of the two galaxies. The mixture of cool and warm/hot gas phases for the absorbing gas and its proximity and orientation to the galaxy pair points to the LLS being a high-velocity gas in the combined halo environment of both galaxies.
\end{abstract}




\section{Introduction}\label{sec:intro}

Circumgalactic medium (CGM) is the gaseous envelope surrounding galaxies, outside their luminous regions but within the virial radii of their dark matter halo, where the gas is still gravitationally bound. Observations of the CGM and the intergalactic gas near galaxies are crucial for understanding the spatial cycling of baryons in and out of galaxies. The CGM has a multiphase structure originating in a number of physical processes operating at galactic scales. The predominant contributors are AGN and star formation-related feedback in the form of supernovae and massive stellar winds, tidal streams resulting from major and minor merger activities, and gas falling into the galaxy's dark matter potential resulting in an envelope of hot corona \citep[e.g.,][]{White91,Veilleux05,Faucher11,Nelson15,Correa18}. The diffuse nature of this multiphased halo gas poses a challenge when attempting to detect it through emission. The more effective approach is to probe the CGM in absorption against the optical and UV continuum of background quasars \citep[e.g.,][]{tumlinson,rudie}.

The gas-rich regions of the CGM manifest as absorption with column densities in the broad range of $\textup{N}(\HI) \approx 10^{16} - 10^{21} \: \cmsq$ \citep{Wotta16}. The lower end of this range includes partial Lyman limit systems (pLLS). The strong {\HI} Damped Lyman Alpha absorbers (DLAs) occupy the higher end, with loosely defined upper and lower limiting values of column densities differentiating these absorber classes \citep[e.g.,][]{Steidel92,Churchill99,Churchill00a,Churchill00b,Rao00}. Multiple studies have revealed a $\gtrsim 3\sigma$ significant inverse correlation between the strength of the {\HI} absorption, measured by the equivalent width and column density, and the projected separation to the nearest galaxy \citep[e.g.,][]{Moller98,Christ07,Monier09,Rao11,Peroux11,Krogager12}. Larger samples do show some scatter in this trend, which is attributed to the uneven distribution of gas surrounding galaxies and the broad range of host galaxies selected by these absorption systems \citep{Rahmati13}. The anti-correlation, despite the scatter, remains evident even at impact parameters of $\rho \gtrsim R_\textup{vir}$, where $R_\textup{vir}$ is the virial radius of the galaxy detected nearest to the absorber. A similar trend of absorption strength with galaxy impact parameter is also seen for some of the strongest metal lines (e.g., {\MgII}, {\CIV}; \citet{Chen00a,Bouche06,Kacprazk08,Chen10,Manuwal21,Churchill13}). Furthermore, correlations are observed between the strength of the absorption lines and other properties of the associated galaxy, such as its luminosity, star formation rate, stellar and halo masses, and redshift \citep{Rahmati14,Nielsen13}. All these establish a strong physical connection between galaxies and the strong {\HI} and metal absorption systems coincident with them. Moreover, as discussed by \citet{Wotta16}, the baryonic overdensities of $\delta \sim 10^2 - 10^3$ corresponding to the typical densities ($n_{\H} \sim 10^{-2} - 10^{-3}$~{\cc}) estimated for the $\log[N(\HI)/\cmsq] \gtrsim 10^{16}$ absorbers is characteristic of circumgalactic gas bound to their central galaxies than gas that is farther away in the intergalactic medium. 

Metallicity estimations of large samples of {\HI} absorption systems have been used to hypothesize on the different origins of the CGM gas. The {\HI} absorption line surveys at $z \leq 1$ find the pLLS and LLS ($16 \lesssim \log[N(\HI)/\cmsq] \lesssim 18$) sampling a wider population of metal-enriched gas of $-3.0 \lesssim \log(Z/Z_{\odot}) \lesssim 0.4$, compared to sub-DLAs and DLAs which are predominantly metal-rich, $\log(Z/Z_{\odot}) \gtrsim -1.4$ \citep{Wotta16,Wotta19}. In other words, the low metallicity absorbers are preferentially probing the lower column density regimes of the CGM. The increase in metallicity with $N(\HI)$ is also seen at $z \gtrsim 2$ \citep{Lehner22b}. When combined with the anti-correlation between the hydrogen column density and the galaxy impact parameter, this suggests the low metallicity absorbers to be located within the CGM but at greater distances from the central galaxy compared to the population of absorbers with higher metallicity. Within the CGM, we expect low metallicities from accreting filaments of metal-poor intergalactic gas and relatively higher metallicities due to outflows and tidally stripped gas from the stellar disks. Indeed, surveys do find such a two-fold distribution in the metallicity range for high {\HI} column density absorbers, with the high and low metallicity sides representing metal-enriched outflows/tidal streams and accretion flows onto galaxies respectively (e.g., \citet{Ribaudo11b,Lehner13,Quiret16,Wotta16,Lehner22b}). Estimates of depletion of elements onto dust in the CGM, when used as a proxy for metallicity, also shows a similar bimodality in the metallicity distribution of absorbers \citep{Wendt21}.

Deeper insights into an absorber's association with the CGM have come from integral field spectroscopic observations of the field centered on the absorbers and also from multi-band photometry of foreground galaxies coupled with stellar population synthesis models. Along with identifying galaxy counterparts to absorbers, the integral field unit (IFU) data cubes provide morphological and kinematic information on the emission line gas, crucial for differentiating between the inflow/outflow phenomena around the galaxy \citep[e.g.,][]{Bouche12,Peroux12,Bouche13,Peroux14,Zabi20,Weng23}. Several such studies have found an excess of absorption along the projected major and minor axes of galaxies, suggesting a non-isotropic distribution of halo gas. Outflowing material appears as preferentially directed along azimuthal angles\footnote{Defined as the angle between the apparent location of the absorber given by the quasar line of sight and the projected major axis of the host galaxy.} within $\approx 30^{\circ}$ of the host galaxy's minor axis, and the accreting gas along the major axis as an extension to the galaxy's disk structure \citep{Bordoloi11,Bouche12,Kacprzak12,Bordoloi14,Schroetter19,Zabl19}. The bimodality in azimuthal angle distribution is more pronounced when the CGM is probed via strong metal lines ({\MgII}, {\OVI}) than {\HI} \citep{Weng23} due to inefficient diffusion of metals within the CGM during the baryon cycle (e.g., \citet{Sameer22}). In comparison, the metallicity of the absorbing gas by itself is less sensitive to the inflow vs. outflow scenario primarily due to the contribution of past outflows to gas accretion at low redshifts \citep{Ford14,Weng23}.

From the absorption line perspective, a complete understanding of the baryonic content within the CGM requires lines that are diagnostic of its complex multiphase structure. In optically thick absorbers, the bulk of the {\HI} and strong low ionization transitions, such as {\MgII}, {\CII}, and {\SiII}, probe the cooler phases of the inflow/outflow gas. Detecting the warm/hot phases requires high ionization lines of species such as {\OVI} and {\NeVIII}. The warm/hot gas dominates the baryon content of the CGM by $\approx 1 - 2$ orders of magnitude compared to the cooler, less ionized gas \citep{Narayanan11,Narayanan12,Narayanan18,Meiring13}. The incidence of {\OVI} in galaxy halos is correlated with the underlying star formation rate in those galaxies and the azimuthal angle at which the line of sight probes the halo \citep{Mason19}. The high ionization gas traced by {\OVI} is found to be ubiquitous (covering fraction $\gtrsim 80\%$) around galaxies where the star-formation rates are high (median SFR $\approx 3$~M$_\odot$~yr$^{-1}$) in contrast with more quiescent systems \citep{Tumlinson11b}. Rather than being uniformly distributed, the {\OVI} associated with such halos is largely confined to regions close to the projected major and minor axes of the central galaxy with covering fractions of $\gtrsim 80\%$ for lines of sight that probe  $\approx 20^{\circ}$ of them. In several instances, the ionization and physical properties of such multiphased gas are reminiscent of the Galactic high and intermediate velocity clouds where the {\OVI} gas phase, with $T \gtrsim 10^5$~K, has its origin in one or more interface layers between cooler CGM gas clouds embedded within the tenuous hot ($T \approx 10^6$~K) corona \citep{Fox06,Narayanan10,Savage10,Savage12}. 

The combined analysis of absorption line data along with 3D spectroscopic information on galaxies coincident with the absorber is a straightforward approach to understanding the physical conditions and metallicities of multiphased gas in the halos of galaxies. In this work, we present the results of such a study involving HST/COS far-UV spectroscopic data of the background quasar FBQS J$0209-0438$, and VLT/MUSE IFU observations of the field surrounding it. Previous studies had reported the presence of a Lyman-limit {\HI} absorber at $\textup{z}_\textup{abs} \approx 0.39$ along this sight line (\citet{Muzahid}: $\log [\textup{N}(\HI)/\textup{cm}^{-2}]$ = 19.00; \citet{Tejos}: $\log [\textup{N}(\HI)/\textup{cm}^{-2}]$ = 18.87; \citet{stevans}: $\log [\textup{N}(\HI)/\textup{cm}^{-2}]$ = 18.00), with no information on the associated metal lines, or the physical properties of the absorber. We present a detailed study of the absorption system and galaxies proximate to it. Throughout
this paper we adopt an $\textup{H}_0$ = 69.6 $\textup{km}\:\textup{s}^{-1}\: \textup{Mpc}^{-1}$ , $\Omega_\textup{M}$ = 0.286, and $\Omega_\Lambda$ = 0.714 cosmology \citep{Bennett14}. 

\section{Spectroscopic Data}\label{sec:spec_data}

The absorption line analysis is performed on the HST/COS spectrum of the QSO (Quasi-stellar Object), FBQS~J$0209-0438$ (RA$=02\textup{h}\: 09\textup{m} \:30.78 \textup{s}$; Dec$=-04^{\circ} 38^{\prime} 26.7^{\prime \prime}$), at an emission redshift of $z = 1.131$ \citep{Veron06}. The data was acquired under Program ID 12264 (Principal Investigator: Simon Morris). The COS observations, obtained with medium spectral resolution gratings (R $\sim$ 20,000), span the wavelength range of $1130 - 1800$ $\textup{\AA}$, with total exposure times of $11.2$ and $22.4$~ks in the G130M and G160M far-ultraviolet gratings, respectively. The co-added spectrum was taken from the HST Spectroscopic Legacy Archive \citep{peeples}. The oversampled spectra were re-sampled to Nyquist criteria of two wavelength pixels per resolution element of $\Delta \lambda = 0.06~\textup{\AA}$. The continuum flux level was defined by fitting lower-order polynomials to $\approx 20$~{\AA} segments, excluding regions with absorption lines. The line measurements were all carried out on the continuum normalized spectrum.

The Keck archive contains HIRES observations at $R \approx 45,000$ of this QSO (Proposal ID: CS280Hr, Principal Investigator: M. Murphy). The HIRES spectra cover the wavelength range of 4018 - 7056~{\AA} with $S/N$ values ranging from $10 - 20$~pixel$^{-1}$. The {\MgIIdblt} lines are outside of the wavelength coverage of HIRES. The optical spectra covers {\CaII}~$3969$, and the weaker lines of {\NaI}, {\MnII}, and {\FeII} which are all non-detections. A system plot showing these significant non-detections in either COS or HIRES spectra is in Appendix, Figure \ref{fig:sys2}.

\begin{figure*}
\begin{center}
\includegraphics[width=0.82\linewidth]{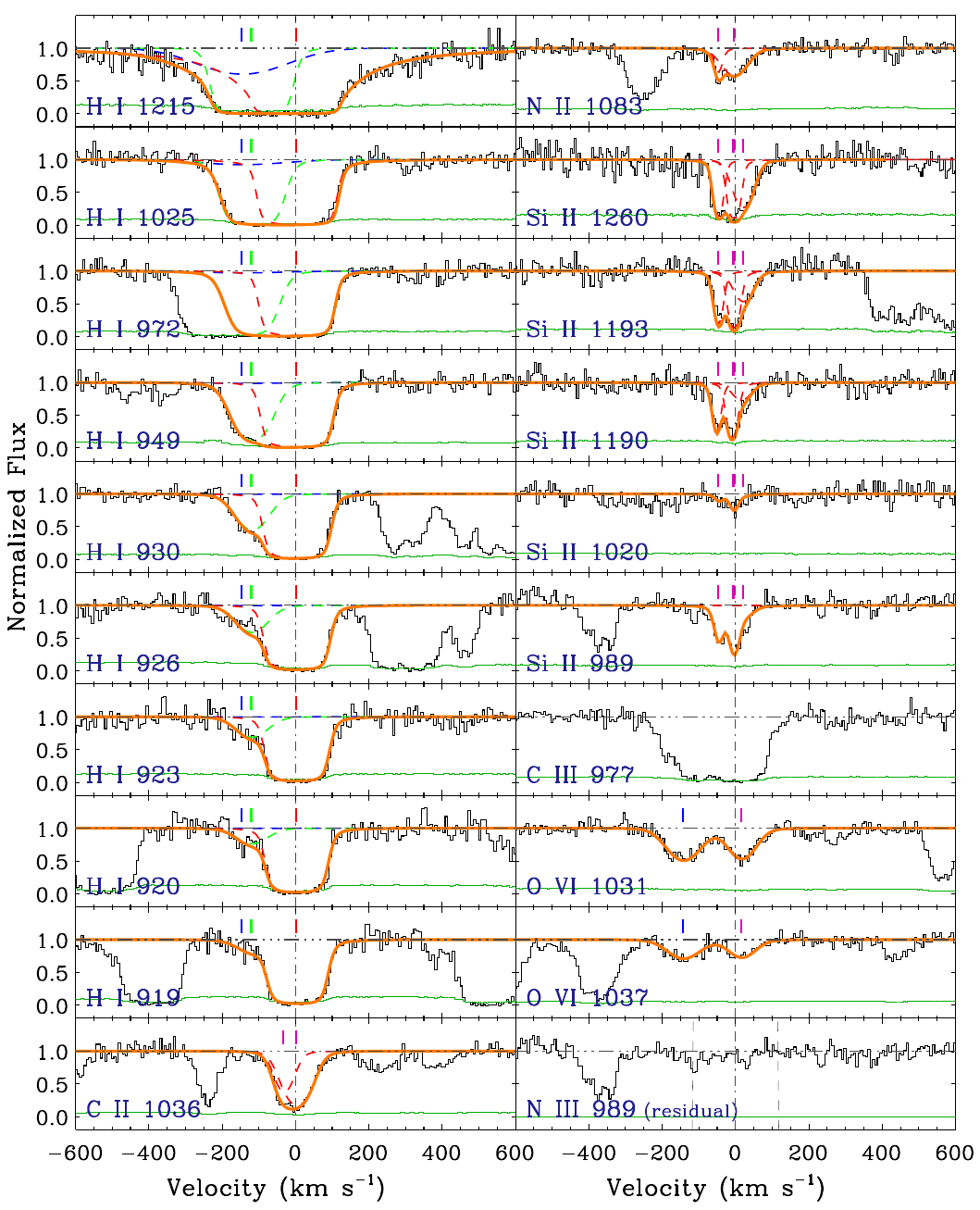}
\caption{The horizontal axis is the velocity in the rest-frame of the absorber, where $v = 0$~{\kms} corresponds to $z = 0.39047$ (vertical \textit{dot-dashed} line). The multi-component cumulative Voigt profile fits are shown in \textit{orange}. The different components are shown separately and their centroids marked by colored vertical markers. The absorption profile of {\CIII} is saturated. The {\NVdblt} lines are covered but not detected. The {\SiIII}~1206 is contaminated. Hence both these lines are shown in Appendix, Figure~\ref{fig:sys2}. For the {\SiII}, we attempt a three-component model based on the profile structure seen for the $1260 - 1190$~{\AA} lines. Such a simultaneous model does not perfectly fit all lines of the multiplet but recovers the total column density for {\SiII} that we obtain from AOD measurements. The column densities of these three components are added together while modeling the LLS component at $v \approx 0$~{\kms}. The column densities of the two components of {\NII} are also added together while modeling the LLS. The {\NIII}~989 panel is the residual flux after subtracting the absorption feature in the {\SiII}~989 panel from its corresponding Voigt profile model. The vertical \textit{dashed} lines in that panel define the velocity range over which the apparent optical depth was integrated to get a lower limit on the {\NIII} column density.}
\label{fig:sys}
\end{center}    
\end{figure*}

\begin{table}
\caption{Equivalent widths of major absorption lines}
\begin{center}
\begin{tabular}{lrr}
\hline
\textbf{Line ID} & \begin{tabular}[c]{@{}c@{}}\textbf{W}\\ \textbf{(m$\boldsymbol{\textup{\AA}}$)}\end{tabular}  & \begin{tabular}[c]{@{}c@{}} \textbf{[$-$v,~+v]} \\ \textbf{($\boldsymbol{\textup{km}\:\textup{s}^{-1}}$)} \end{tabular} \\ \hline \hline
{\HI} 1215	&	1753 $\pm$ 24 	&	$[-345, 273]$	\\
{\HI} 1025	&	1110 $\pm$ 10  	&	$[-275, 185]$	\\
{\HI} 949	&	891 $\pm$ 11  	&	$[-275, 185]$	\\
{\HI} 930	&	711 $\pm$ 10  	&	$[-275, 185]$	\\
{\HI} 926	&	671 $\pm$ 15  	&	$[-235, 150]$	\\
{\HI} 923	&	635 $\pm$ 16 	&	$[-235, 150]$	\\
{\HI} 920	&	592 $\pm$ 18	&	$[-235, 150]$	\\
{\HI} 919	&	531 $\pm$ 13 	&	$[-140, 115]$	\\
{\HI} 918	&	526 $\pm$ 13 	&	$[-140, 115]$	\\
{\HI} 917	&	495 $\pm$ 10  	&	$[-106, 113]$	\\
{\CII} 1036	&	365 $\pm$ 8  	&	$[-118, 115]$	\\
{\NII} 1083	&   183 $\pm$ 10  	&	$[-118, 115]$	\\
{\NIII} 989	&   $\gtrsim$ 52  	&	$[-118, 115]$	\\
{\OVI} 1031	&	346 $\pm$ 12  	&	$[-235, 105]$	\\
{\OVI} 1037	&	169 $\pm$ 11  	&	$[-235, 105]$	\\
{\SiII} 1260	&	459 $\pm$ 19 	&	$[-118, 115]$	\\
{\SiII} 1193	&	319 $\pm$ 15 	&	$[-118, 115]$	\\
{\SiII} 1190	&	281 $\pm$ 15 	&	$[-118, 115]$	\\
{\SiII} 1020	&	44 $\pm$ 13 	&	$[-118, 115]$	\\
{\SiII} 989	    &	233 $\pm$ 11  	&	$[-118, 115]$	\\
{{\NV} 1242}    &   {< 87}          & {$[-235, 105]$} \\
{{\CaII} 3969}    &   {< 92}            &   {$[-118, 115]$} \\
{{\FeII} 1144}    &   {< 107}           & {$[-118, 115]$} \\
{{\FeIII} 1122}   &   {< 66}         &   {$[-118, 115]$} \\ 
\hline
\end{tabular}
\end{center}
\tablecomments{{Equivalent widths in the rest-frame of the absorber for lines detected at $\geq 3\sigma$ significance, and the corresponding velocity range used for measuring the equivalent width. The expected location of {\NIII}~$989$ overlaps with the {\SiII}~$989$ feature. The listed value is from integrating the residual obtained after removing the estimated {\SiII}~$989$ absorption at that velocity as explained in Section \ref{sec:spec_data}. The {\HI}~972, {\HI}~937, and {\SiIII}~1206 are not included as they are contaminated. The {\CIII}~977 is not included because of its strong saturation. The expected location of {\FeII}~1144 line is partially contaminated by an absorption feature that is offset by $-41$~{\kms} resulting in a higher upper limit for the non-detection. \citet{Tejos} identify the contaminant as {\Lya} at $z = 0.31$. The {\CaII}~3969 limit is from the Keck/HIRES data.}}
\label{tab:ew}
\end{table}

Voigt profile models were fitted to the absorption lines using the VPFIT\footnote{https://people.ast.cam.ac.uk/$\sim$rfc/vpfit.html} routine (ver. 10.4) \citep{Carswell14} to determine the column densities, Doppler $b$-parameters, and velocity centroids of the absorption features. The profile models were convolved with the COS line-spread functions for the nearest corresponding wavelength in the observed spectrum. Absorption profiles created by different transitions of a species were fit together to obtain unique solutions for the column densities and Doppler $b$-parameters. Total column densities for the detected ions were also measured by integrating the pixel-by-pixel apparent optical depth (AOD) across the absorption feature following the technique given by \citet{savage}. For lines that are non-detections, an upper limit on the column density was arrived at from the 3$\sigma$ upper limit on the equivalent width, assuming the linear part of the curve of growth.

\begin{figure*}
\begin{center}
    \includegraphics[width=\linewidth]{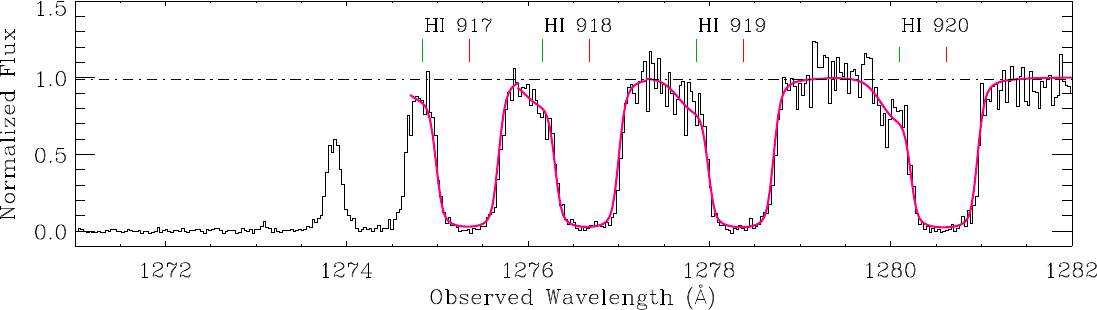}
    \caption{This plot shows the higher order Lyman series lines, with the flux shown in black; the mean continuum level is shown by the horizontal dot-dashed line; the magenta curves are the two-component voigt profile fits to the Lyman lines with the first and second components marked by the red and green markers, respectively. A very high neutral hydrogen ``column'' density causes a sharp decline in the flux around the location of the redshifted ($z \sim 0.39$) Lyman limit break.}
    \label{fig:optical}
  \end{center}    
\end{figure*}

\begin{figure*}
\begin{center}
\includegraphics[scale=0.8]{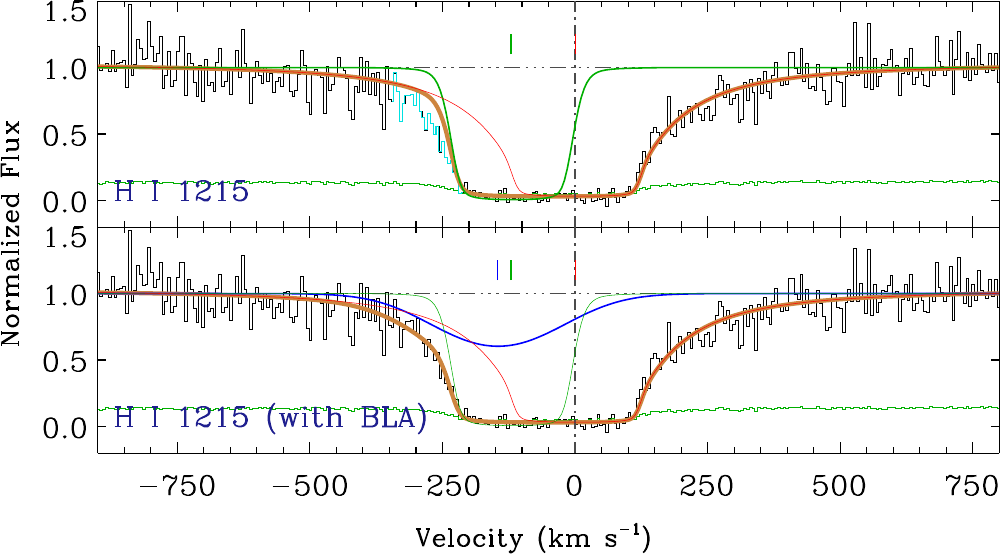}
\caption{Voigt profile fits for the Ly$\alpha$ transition. The top panel shows the two-component voigt profile fit to the absorption line, with the first and second-component fits shown in red and green colors, respectively, and the resultant combined fit is shown by the thick profile. The pixels in cyan color are not fit properly by the current voigt profile. The bottom panel shows a three-component voigt profile fit (the thick profile), which beautifully fits the entire absorption line. The first, second, and third component fits are shown in the colors red, green, and blue, respectively. The third component is a Broad Lyman Alpha (BLA), as is clear from the broad and shallow profile.}
\label{fig:bla_fit}
\end{center}
\end{figure*}

\begin{table*}
\begin{center}
\caption{Results from Voigt Profile Fitting \& Integrated AOD Method}
    \begin{tabular}{l r r r} \hline

\begin{tabular}[c]{c@{}}\textbf{Line ID}  \end{tabular} & \begin{tabular}[c]{c@{}}\textbf{v (in $\boldsymbol{\textup{km}\: \textup{s}^{-1}}$)}  \end{tabular}  & \begin{tabular}[c]{c@{}}\textbf{b (in $\boldsymbol{\textup{km}\: \textup{s}^{-1}}$)}  \end{tabular} & \begin{tabular}[c]{c@{}}\textbf{$\boldsymbol{\log [\textup{N}/\textup{cm}^{-2}]}$}  \end{tabular} \\ \hline \hline
\begin{tabular}[c]{@{}c@{}}{\HI} 1215/1025/949/930/\\ 926/923/920/919/918/917/916\end{tabular} & \begin{tabular}[c]{@{}c@{}c@{}}$-2^\textup{a} \pm 2$\\ $-121^\textup{b} \pm 3$ \\ $-147^\textup{c} \pm 3$\end{tabular} & \begin{tabular}[c]{@{}c@{}c@{}}$34 \pm 3$\\ $54 \pm 3$ \\ $156 \pm 30$\end{tabular} & \begin{tabular}[c]{@{}c@{}c@{}}$18.65 \pm 0.20$\\ $15.86 \pm 0.10$ \\ $14.01 \pm 0.20$\end{tabular} \\ \hline

\begin{tabular}[c]{c@{}}{\OVI} 1031/1037 \end{tabular}& \begin{tabular}[c]{@{}c@{}}$20^\textup{a} \pm 3$\\ $-139^\textup{b,c} \pm 3$\end{tabular}  & \begin{tabular}[c]{@{}c@{}}$31 \pm 4$\\ $42 \pm 5$\end{tabular} & \begin{tabular}[c]{@{}c@{}}$14.24 \pm 0.05$\\ $14.33 \pm 0.05$\end{tabular} \\ \hline

\begin{tabular}[c]{c@{}}{\CII} 1036 \end{tabular} & \begin{tabular}[c]{@{}c@{}}$-33^\textup{a} \pm 3$\\ $2^\textup{a} \pm 4$\end{tabular}  & \begin{tabular}[c]{@{}c@{}}$30 \pm 2$\\ $41 \pm 3$\end{tabular} & \begin{tabular}[c]{@{}c@{}}$14.30 \pm 0.05$\\ $14.69 \pm 0.04$\end{tabular} \\ \hline

\begin{tabular}[c]{c@{}}{\SiII} 1260/1193/1190/1020/989 \end{tabular} & \begin{tabular}[c]{@{}c@{}c@{}}$-45^\textup{a} \pm 2$\\ $-6^\textup{a} \pm 3$ \\ $19^\textup{a} \pm 2$\end{tabular} & \begin{tabular}[c]{@{}c@{}c@{}}$9 \pm 3$\\ $11 \pm 2$ \\ $31 \pm 3$\end{tabular} & \begin{tabular}[c]{@{}c@{}c@{}}$14.04 \pm 0.09$\\ $14.37 \pm 0.10$ \\ $13.36 \pm 0.06$\end{tabular} \\ \hline

\begin{tabular}[c]{c@{}}{{\NII} 1083} \end{tabular} & \begin{tabular}[c]{@{}c@{}}$-48^\textup{a} \pm 4$\\ $-5^\textup{a} \pm 4$\end{tabular}  & \begin{tabular}[c]{@{}c@{}}$10 \pm 2$\\ $42 \pm 2$\end{tabular} & \begin{tabular}[c]{@{}c@{}}$13.77 \pm 0.12$\\ $14.20 \pm 0.03$\end{tabular} \\ \hline

\begin{tabular}[c]{c@{}}{{\FeIII} 1122} \end{tabular} & \begin{tabular}[c]{c@{}}$-$ \end{tabular}  & \begin{tabular}[c]{c@{}}$-$ \end{tabular}   &  \begin{tabular}[c]{c@{}}${< 14.1}$ \end{tabular}  \\ \hline

\begin{tabular}[c]{c@{}}{{\FeII} 1144} \end{tabular} & \begin{tabular}[c]{c@{}}$-$ \end{tabular}  & \begin{tabular}[c]{c@{}}$-$ \end{tabular}   &  \begin{tabular}[c]{c@{}}${< 14.0}$ \end{tabular}  \\ \hline

\begin{tabular}[c]{c@{}}{\NIII} 989 \end{tabular} & \begin{tabular}[c]{c@{}}$-$ \end{tabular}  & \begin{tabular}[c]{c@{}}$-$ \end{tabular}   &  \begin{tabular}[c]{c@{}}${\geq 13.8}$ \end{tabular}  \\ \hline        

\begin{tabular}[c]{c@{}}{\NV} 1242 \end{tabular} & \begin{tabular}[c]{c@{}}$-$ \end{tabular} & \begin{tabular}[c]{c@{}}$-$ \end{tabular} &  \begin{tabular}[c]{c@{}}$< 13.9$ \end{tabular} \\ \hline

\begin{tabular}[c]{c@{}}{\CIII} 977 \end{tabular} & \begin{tabular}[c]{c@{}}$-$ \end{tabular} & \begin{tabular}[c]{c@{}}$-$ \end{tabular} &  \begin{tabular}[c]{c@{}}$> 14.70$ \end{tabular} \\ \hline

\end{tabular}
\end{center}
\flushleft{\tablecomments{The velocity centroids of the components, their Doppler $b$-parameters, and column densities obtained from Voigt profile fitting are listed in the different columns. For {\CII}, {\NII}, and {\SiII} the column density we adopt while modeling the absorption at $v(\HI) \approx 0$~{\kms} ($z \sim 0.39047$) is the total column density of the individual components added together. {For the {\NV}~$1242$, {\FeII}~$1144$, and {\FeIII}~$1122$ lines, which are non-detections, the upper limit on column density is obtained by integrating the apparent column density over the same velocity range as lines of similar ionization. The {\CIII} column density is a lower limit due to its strong saturation. The limit was obtained by integrating the AOD across the profile from $[-260, 140]$~{\kms}. And the lower limit on {\NIII} column density is obtained by integrating the AOD across the residual shown in Figure \ref{fig:sys} from $[-118,115]$~{\kms}. The superscripts in the velocity column denote the components of the absorber that the various voigt profile fits would contribute to: a$-$first component (LLS in PIE); b$-$second component (lower density cloud in PIE); c$-$third component (collisionally ionized BLA).}}}
\label{tab:vpfit_res}
\end{table*}

\section{IFU Data}\label{sec:ifu}

The quasar field was observed through the integral field spectrograph $VLT$/MUSE \citep{Bacon17} operating in wide-field mode, capturing a 1 arcmin $\times$ 1 arcmin field of view. The spectral coverage spans $4750 - 9351\: \textup{\AA}$ and was part of the MUSEQuBES (MUSE Quasarfield Blind Emitters Survey) \citep{Dutta23}, which aimed to analyze low-redshift galaxies associated with 16 UV-bright QSOs. For this specific quasar, observations were conducted over a total of 8 hours (P.I: Joop Schaye) under program IDs $094.\textup{A}-0131, 095.\textup{A}-0200, 096.\textup{A}-0222, 097.\textup{A}-0089, \textup{and} \: 099.\textup{A}-0159$. Each $1-$hour observation block was divided into four $900-$second exposures, with a $90^{\circ}$ rotation and offset of a few arcsecs. Data reduction employed both the MUSE data reduction pipeline (v1.2) and procedures from the \texttt{CubeX} package.

The final reduced data cube consists of 433 $\times$ 433 spatial pixels (spaxels)\footnote{Spatial pixels in the IFU data cube containing associated spectra}, with each spaxel comprising 3682 pixels along the spectral axis. The spectral range spans from $4750$ to $9350 \: \textup{\AA}$, 
with the spectral resolution varying from R$\sim$1800 at $\lambda = 5000 \: \textup{\AA}$ to R$\sim$3500 at  $\lambda =8000 \: \textup{\AA}$. The spatial sampling employs a grid of $0.2^{\prime \prime} \times 0.2^{\prime \prime} $ pixels.

The central 20 spaxels within the IFU data cube are dominated by the quasar point spread function (PSF). To explore close line-of-sight separations resulting in GOTOQ configuration (Galaxy On Top Of Quasar), the quasar continuum was subtracted from the data cube. The first step to this was to estimate the parameters of quasar PSF, which was done using the PampelMuse\footnote{https://pampelmuse.readthedocs.io/en/latest/} software for a 2D Moffat function that is used to model the quasar PSF. The parameters are modeled for the entire wavelength range for multiple stars, and their values, FWHM $= 0.563$ arcsecond and $\beta = 2.07$, are obtained at $\lambda = 8000$~{\AA} for which the PSF profile aligns for all stars. After obtaining the PSF parameters, the quasar light was subtracted using \citet{Johnson18}'s algorithm.

\section{Spectroscopic analysis of the absorption system}\label{sec:spec_analysis}

In Figure \ref{fig:sys}, we show the {\HI} Lyman series and the metal lines detected in the absorber’s rest frame. Absorption from {\CII}, {\SiII}, {\CIII}, {{\NII}}, {\NIII}, and {\OVI} are detected at $> 3\sigma$ significance. The equivalent width measurements are listed in Table \ref{tab:ew}. The {\NVdblt} {along with {\FeII} 1144 and {\FeIII} 1122 are prominent non-detections. The {\SiIII}~1206 line is contaminated, as described in the Appendix (see Figure \ref{fig:sys2})}, making it useless for column density measurement. Table \ref{tab:vpfit_res} lists the results from voigt profile fitting.

In the Lyman series lines, from {\HI}~949~{\AA} and above, the presence of at least two components at $v \approx 0$~{\kms} and $v \approx -121$~{\kms} are seen, with the latter weaker in strength. A two-component Voigt profile model yields $\log [\textup{N}(\HI)/\textup{cm}^{-2}] = 18.65$ and $15.86$ for the two components, respectively. The $v \approx 0$~{\kms} component is responsible for the Lyman break at $1273$~{\AA} (see Figure \ref{fig:optical}). The column density obtained from profile fitting is consistent with the lower limit of $\log [\textup{N}(\HI)/\textup{cm}^{-2}] > 18$ given by the optical depth at the Lyman break ($\boldsymbol{{\tau}_{912\:\textup{\AA}} > 7.1}$). The {\CIII}~$977$~{\AA} shows saturated absorption at both these velocities. The low ionization metal lines ({\SiII}, {\CII}, {{\NII}}, and {\NIII}) are, however, detected only at the velocity of the Lyman break component. 

The {\SiII},  {\CII}, {and {\NII} lines} show further sub-component structure to the absorption at $v \approx 0$~{\kms} with two components separated by $\Delta v \approx 40$~{\kms}. There is a possible weaker third component at $v \approx +20$~{\kms} that is evident only in the stronger members of the {\SiII} multiplets. We, therefore, fit a three-component profile simultaneously to the {\SiII} lines. A component structure corresponding to this cannot be discerned in the higher-order Lyman series lines because of saturation, restricting component-by-component modeling. When added together, the individual column densities from a two or three-component model for the low ions are within $1\sigma$ of the column density obtained using a single-component profile model. For modeling, we consider the column densities of these blended components at $v \approx 0$~{\kms} together for {\CII}, {\NII}, and {\SiII}. 

{The expected location of the {\NIII}~989 ($\lambda_0 = 989.799$~{\AA}, $f_{osc} = 0.123$) overlaps with the {\SiII}~$989$ ($\lambda_0 = 989.8731$, $f_{osc} = 0.171$). The close correspondence between the observed profile and the uncontaminated {\SiII}~$1260, 1193, 1190$ lines suggest that a significant portion of the absorption at $\lambda = 1376.3$~{\AA} likely originates from {\SiII}~$989$. Fitting the four uncontaminated {\SiII} lines (including {\SiII}~1020) simultaneously yields nearly identical {\SiII} column density values compared to using all five lines. To determine a lower limit on the $N(\NIII)$, we subtract from the observed profile the expected {\SiII}~$989$ absorption derived from a simultaneous fit to the {\SiII} lines (see Figure \ref{fig:sys}). By integrating the residual thus obtained over the same velocity range as the {\CII}, and {\SiII}, we obtain $\log~N_a (\NIII) = 13.80~\pm~0.10$, which we adopt as a lower limit on the {\NIII} column density as the true contribution of {\NIII} to the absorption is likely higher than this.}

The {\OVIdblt} lines have two kinematically distinct unsaturated components. Simultaneous Voigt profiles for the doublet lines give component centroids at $v \approx 20$~{\kms} and $v \approx -139$~{\kms}, which are different from the velocity centroids of the low and intermediate metal ions, and the two distinct components in {\HI}. Differences in the kinematic profiles of {\OVI} with {\CII}, {\SiII}, and the core absorption in {\HI} is the signature of multiphase gas (e.g., \citet{Fox13,Kacprzak19}). 

The two-component fit to {\HI} absorption is primarily governed by the higher-order Lyman lines where the component structure is evident. The component responsible for the Lyman break is less saturated at higher orders, and the offset component is unsaturated. Such a model does not, however, fit the excess absorption in the blue wing of the {\Lya} in the velocity range of [$-420, -240$]~{\kms} (see Figure \ref{fig:bla_fit}). The column density of the nearest ($v \approx -121$~{\kms}) {\HI} component is strongly constrained by its unsaturated profile and non-detection in the highest orders. The excess absorption in the {\Lya} indicates the possible presence of an additional component that could be shallow and absent in the weaker {\HI} transitions. A third component in {\HI} is also consistent with the presence of {\OVI} at a velocity that is different from the low ionization metal lines. 

A fit to the {\HI} with such a third component centered around the velocity of the nearest {\OVI} ($v\approx-147$~{\kms}) is shown in Figure \ref{fig:bla_fit}, with best-fit values of $\log [\textup{N}(\HI)/\textup{cm}^{-2}] = 14.02\pm0.09$, and $b(\H) = 155\pm3$~{\kms}. The statistical uncertainties given by the fitting routine are an underestimation. The fit parameters for this third component are sensitive to the velocity of the second component. Changing the velocity centroid of the second {\HI} component between its $1\sigma$ limits of $v = -118$~{\kms}, and $v = -124$~{\kms}, we obtain best-fit values for the broad {\Lya} component as $\log [\textup{N}(\HI)/\textup{cm}^{-2}] = 14.01^{+0.20}_{-0.11}$, and $b(\H) = 156^{+24}_{-38}$~{\kms}. This adopted error in $b$ could be more if the component structure in {\HI} is more complex than the assumed three components. The continuum in the region of the {\Lya} line is reasonably well defined (see Figure \ref{fig:bla_fit}). The placement of the continuum is, therefore, unlikely to be a major source of uncertainty in the BLA fit parameters. 

\begin{figure*}
\begin{center}
\begin{minipage}{0.40\textwidth}
    \centering
\includegraphics[width=\linewidth]{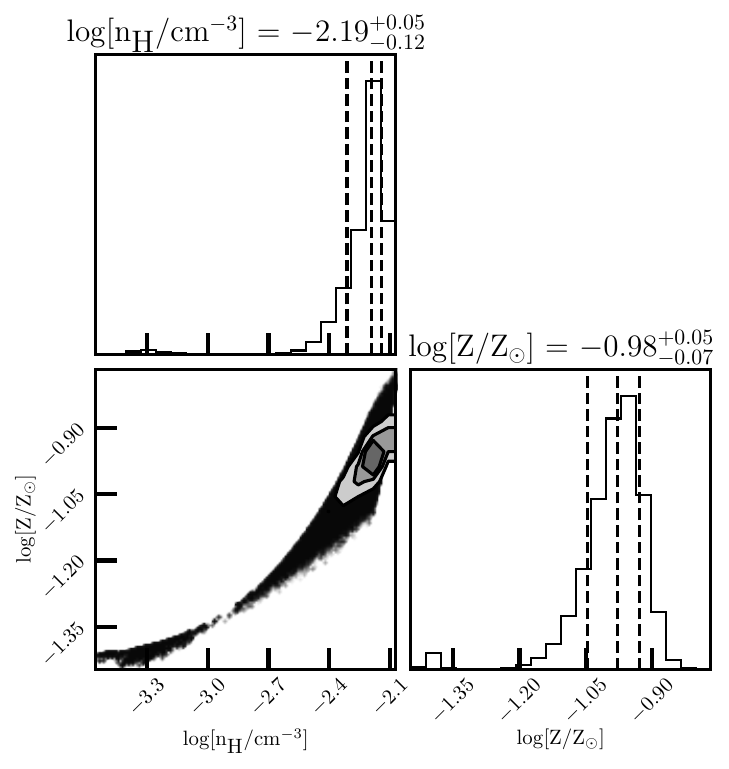}
\end{minipage}
\begin{minipage}{0.59\textwidth}
    \centering
\includegraphics[width=\linewidth]{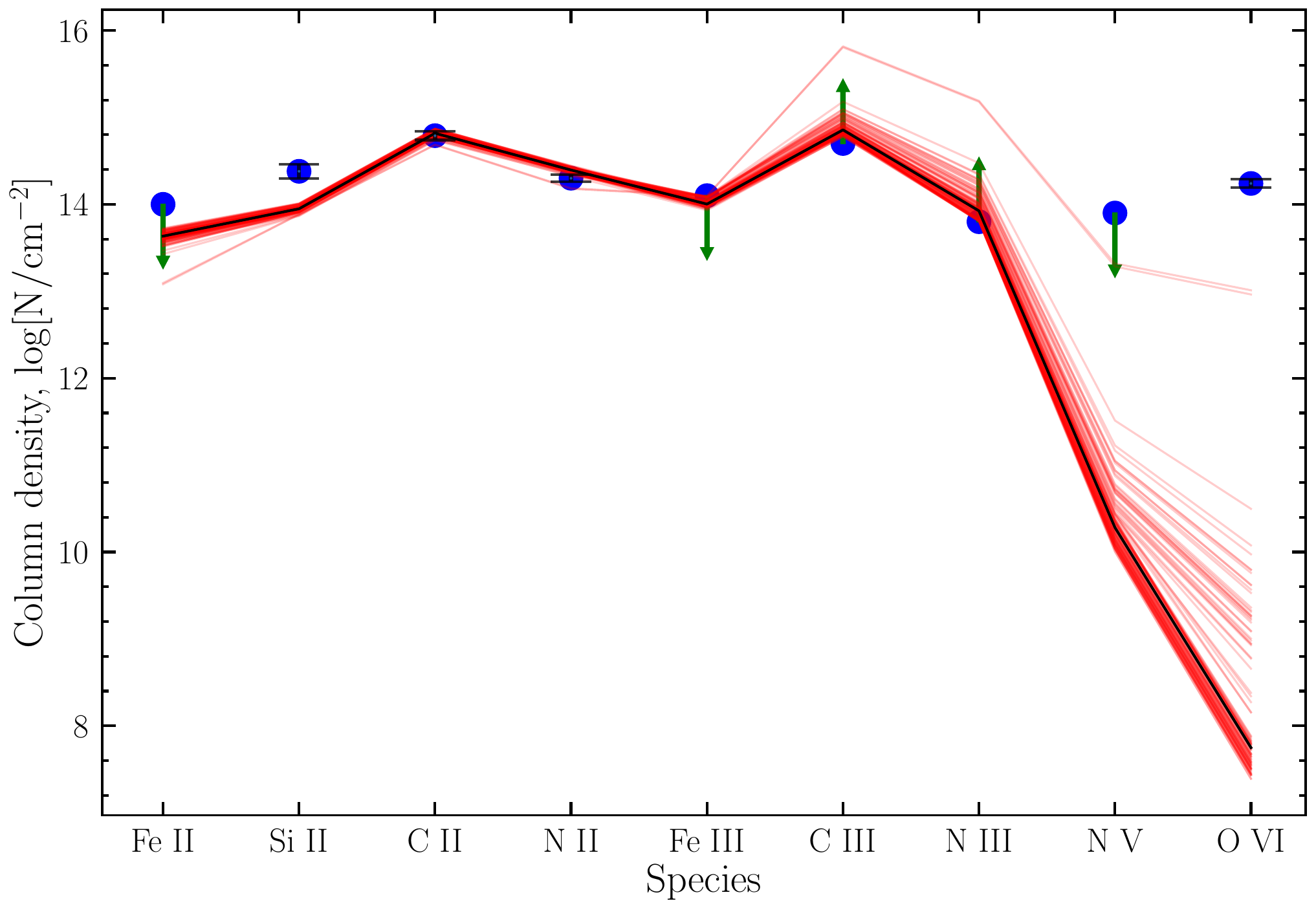}
\end{minipage}
\caption{\textbf{Left:} The corner plot from photoionization modeling of the 1st component responsible for the Lyman break. It shows the posterior and marginalized distribution of the absorber's metallicity (Z) and hydrogen number density ($n_{\H}$). The values denoted by dashed lines are the median and 16-84 percentile values of the marginalized distribution. \textbf{Right:} This plot shows the observed (blue) column densities of the species {\CII}, {\SiII}, {{\NII}}, and {\OVI} in the increasing order of their ionization potentials, with 1$\sigma$ error bars. For {{\FeII}, {\FeIII}, {\NV}, {\NIII}, and {\CIII}}, we have indicated the observed upper and lower limits with arrows. The red curves show the column density predictions for random samples of $n_{\H}$ and Z from posteriors. The thick black line shows the column density predictions for the median values of $n_{\H}$ and Z. It can be seen that {\SiII} is under-predicted by $\sim 0.4$ dex, suggesting a sub-solar [C/$\alpha$] value. For other species, the predictions are in agreement with the observations, except for {\OVI}, which implies that {\OVI} exists in a different phase than the rest of the species detected in this component.}
\label{fig:data_model}
\end{center}    
 \end{figure*}

\section{Photoionization Modeling}\label{sec:modeling}

We run a suite of {\CLOUDY} (ver C17.01, \citet{Ferland13}) models to determine the physical conditions and chemical abundances of the absorbers. These models assume that absorbing clouds are plane parallel slabs with uniform density and metallicity. 
The ionization state of each ion is calculated under the assumption that the clouds are in thermal and photoionization balance, regulated by the extragalactic UV background. For 
this, we used the UV background model by \citet[][their fiducial Q18 model]{Khaire19}, which incorporates updated emissivities of quasars and galaxies 
\citep[from][]{Khaire15puc, Khaire15ebl, Khaire16}. We modeled the three kinematically distinct components separately.

We adopt the Bayesian MCMC (Monte Carlo Markov Chain) modeling approach described in \citep{Acharya22} for the component-wise multiphase modeling of the absorption system. The measured {\HI} column density as an input parameter is used as the stopping criteria in the models. For the chosen hydrogen number density, $n_{\H}$, the model increments the path length through the absorbing cloud until the measured {\HI} column density is recovered. The metal ion column densities are then computed for this model cloud, assuming a relative chemical abundance pattern similar to solar as given by \citet{Grevesse}. 

For the component at $v \approx 0$~{\kms} responsible for the Lyman break, the model is unable to converge to a solution that simultaneously explains {\CII}, {\SiII}, {{\NII}}, and {\OVI}, {along with predictions that are consistent with the limits on {\NIII}, {\NV}, {\FeII}, {\FeIII} and {\CIII}}. {Including all these species as constraining ions results in the Bayesian MCMC failing to converge onto a solution.} On the other hand, if we remove {\OVI} from the list of constraining ions, a single-phase solution, as shown in Figure \ref{fig:data_model}, is obtained. The values for the density and metallicity of this phase are {$n(\H) = (4.9 - 7.2) \times 10^{-3}$~{\cc}} and {$\log~[Z/Z_\odot] = (-1.1, -0.9)$}, respectively, where the ranges correspond to $1\sigma$ deviations from the maximum likelihood value. {The predicted {\SiII} column density for these models is $\approx 0.4$~dex lower than the observed value, suggesting a sub-solar abundance pattern. \citet{Zahedy21} found that LLSs exhibit a wide range of elemental abundances ($-1.0 \lesssim [\textup{C}/\alpha] \lesssim +0.7$) reflecting their enrichment history, with a slight trend of decreasing [C/$\alpha$] with decreasing metallicity. The $[\textup{C}/\alpha]\approx-0.4$ which we find is consistent with such a trend.} The model results are summarized in Table \ref{tab:phases}. As shown in Figure \ref{fig:data_model}, the best model predicted column densities of {\OVI} are {approximately six orders of magnitude} smaller than the observed column density, suggesting that {\OVI}, though kinematically coincident with the Lyman limit component, is tracing a separate gas phase. 

\begin{figure}
\begin{center}
 \includegraphics[width=\columnwidth]{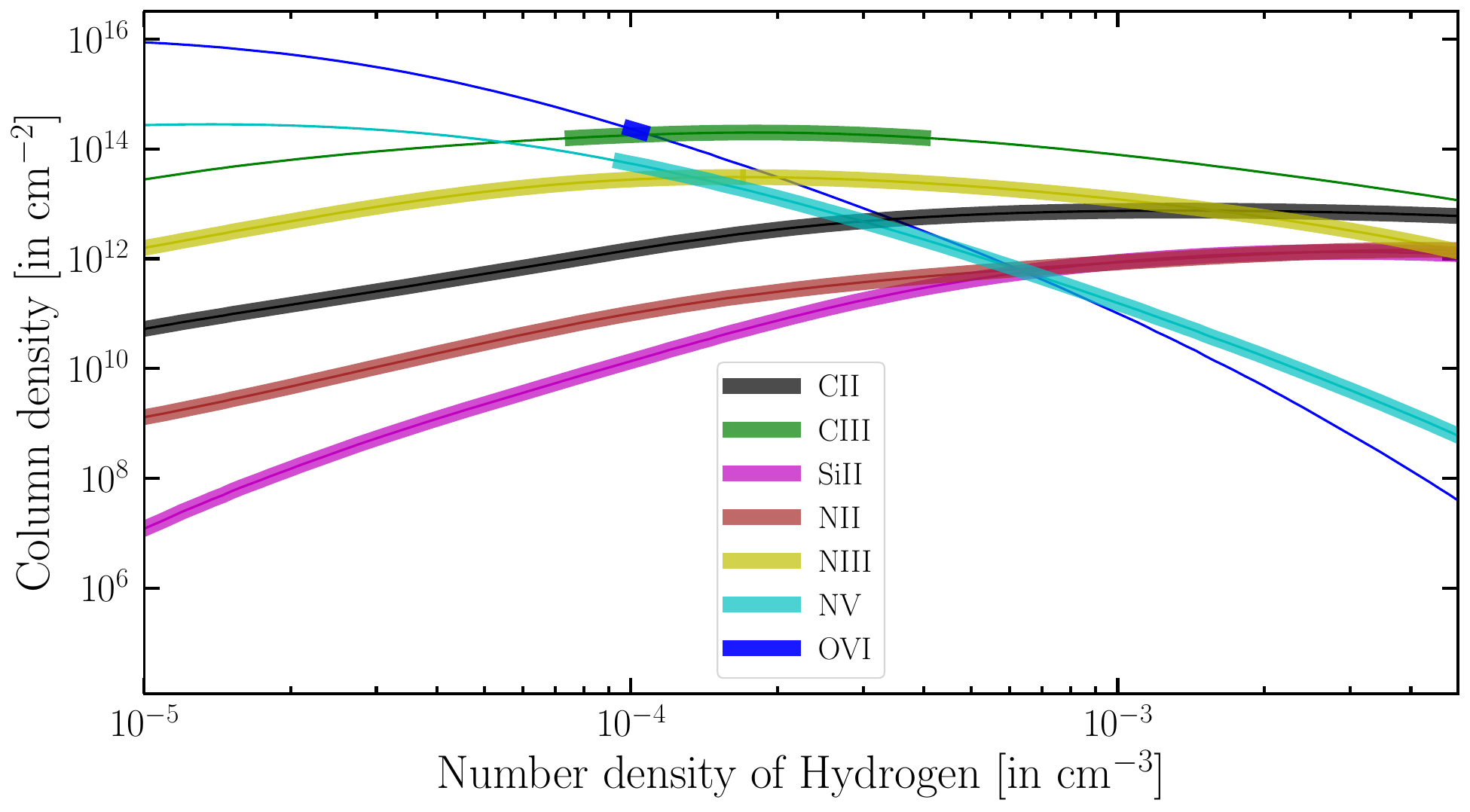}
\caption{Column densities of the {\CII}, {\CIII}, {\SiII}, {\NIII}, {\NV}, {\OVI} predicted by {\CLOUDY} with respect to the number density of hydrogen for metallicity, $\log (\textup{Z}/\textup{Z}_{\odot})=-0.90$, which is a lower limit. The highlighted portions of the curves show the range of hydrogen number density values that satisfy the observations. The saturated {\CIII} absorption is not explained if the metallicity goes below the lower limit. {Also, the lower limit on {\NV} column density due to non-detection is explained only if the metallicity of Nitrogen is 0.2 dex lower than the rest of the elements.} Thus, we obtain a range for hydrogen number density for which all the observations can be explained.}
\label{fig:comp2}
 \end{center}    
\end{figure}

The only metal ions with absorption near the velocity of the second component of {\HI} at $v \approx -121$~{\kms} are {\CIII} and {\OVI}. All other metal ions are non-detections. The centroid of {\OVI} is offset from the {\HI} by $\sim 20$~{\kms}, which is about one resolution element of COS. The absorption in {\CIII} is saturated and provides us with only a lower limit on the column density. Since there are not enough constraining ions for the MCMC modeling to converge to a solution, we plot the {\CLOUDY} predictions as shown in Figure \ref{fig:comp2} with limits on metallicity and density for this component. {We find that for $\log~[Z/Z_{\odot}] \lesssim -0.9$, the saturated {\CIII} is not recovered at any density by the photoionization equilibrium (PIE) models. Adopting this lower limit of $-0.9$~dex for metallicity the nearest observed {\OVI} for this component is recovered by the PIE models (see Figure \ref{fig:comp2})  at $n_{\H} = 1 \times 10^{-4}$~{\cc} with a corresponding line of sight thickness of $L = 187$~kpc, which is quite large. For metallicity larger than $-0.9$ dex, the {\OVI} and {\CIII} can be simultaneously recovered at $n_{\H} \gtrsim 10^{-4}$~{\cc}, and correspondingly lower path lengths. The PIE models thus predict a lower limit on the metallicity for the $v \approx -121$~{\kms} which is comparable to the metallicity of the central LLS to within its $1\sigma$. Differences in metallicities between components of an absorber are not unusual. Large {\HI} column density components tend to have comparatively lower metallicities due to the inefficient mixing of gas and metals on small scales \citep{Schaye07,Sankar20}.} In these acceptable models, the abundance of Nitrogen has to be lower than {$0.2$~dex} for the {\NV} to remain a non-detection. The limits on the properties of this phase, obtained by running a {\CLOUDY} photoionization suite with $\log~[Z/Z_{\odot}] \approx -0.9$ and $\textup{n}_\textup{H} \approx 10^{-4}$~{\cc}, are summarized in Table \ref{tab:phases}. 
 
\begin{table*}
\renewcommand{\arraystretch}{1.5}
\centering
\caption{Ionization Modeling Results}
\begin{tabular}{l r r c} \hline
\textbf{Parameters}             & \textbf{Comp 1 (LLS, PIE)}   & \textbf{Comp 2 (PIE)}                         & \textbf{Comp 3 (BLA, CIE)}                 \\
\hline \hline
\textbf{Constraining Ions} & {\CII}, {\SiII}, {{\NII}, {\NIII}, {\NV}, {\FeIII}, {\FeII}, {\CIII}} & {{\NV}}, {\OVI}, {\CIII} & {\OVI} \\ 
\textbf{v (in $\boldsymbol{\textup{km}\: \textup{s}^{-1}}$)}          & $-$2                      & $-$121                                      & $-$147                                      \\
\textbf{T (in K)}          & {$1.37 \times 10^4$}                & $\lesssim 2.67 \times 10^4$                                 &  $1.38^{+0.63}_{-0.62} \times 10^6$                       \\
\textbf{P/k (in $\textup{cm}^{-3}\:\textup{K}$)}          & {$192$}                & $\gtrsim 6.02$                                 &  $-$ \\
\textbf{L (in kpc)}        & {$2.43$}                   & $\lesssim 187$                                    & $-$                                   \\
\textbf{$\textup{n}_\textup{H}$ (in $\textup{cm}^{-3}$)}      & {$6.46 \times 10^{-3}$}                & $\gtrsim 1 \times 10^{-4}$                                  & $-$                                  \\
\textbf{log (Z/$\textup{Z}_{\odot}$)}      & {$-0.98^{\ast} \pm 0.07$}            & $\gtrsim -0.90$                                         & $\sim -0.50$                                         \\
\textbf{N({\HI}) (in $\textup{cm}^{-2}$)}  & $4.47 \times 10^{18}$                & $7.24 \times 10^{15}$                                  & $1.02 \times 10^{14}$     \\
\textbf{N(H) (in $\textup{cm}^{-2}$)}  & {$4.85 \times 10^{19}$}                & $\gtrsim 5.77\times 10^{19}$                                  & ${6.85 \times 10^{20}}$     \\\hline
\end{tabular}
\flushleft{\tablecomments{The results for the three components observed in the absorber system located at a redshift of 0.39047. The first two components of this complex are photoionized, with a PIE temperature of $\sim 10^4$ K. This is the cooler phase with two distinct kinematic components of different number densities and metallicities, and together containing baryons with column density $\sim 10^{20}\:{\cmsq}$. Whereas the third component, traced by the BLA and {\OVI}, is collisionally ionized with a CIE temperature of $\sim 10^6$ K. It has very low neutral hydrogen column density. Still, it contains {almost an order of magnitude more baryons} ($\textup{N(H)}\sim 6.85\times10^{20}$ \cmsq) than the cooler phase.\\ {*The metallicity shown here is following a solar abundance pattern, but there are deviations from the solar abundance, e.g.: the relative abundance of Si is higher than in the sun.}}}
\label{tab:phases}
\end{table*}

\begin{figure}
\begin{center}
\includegraphics[width=\columnwidth]{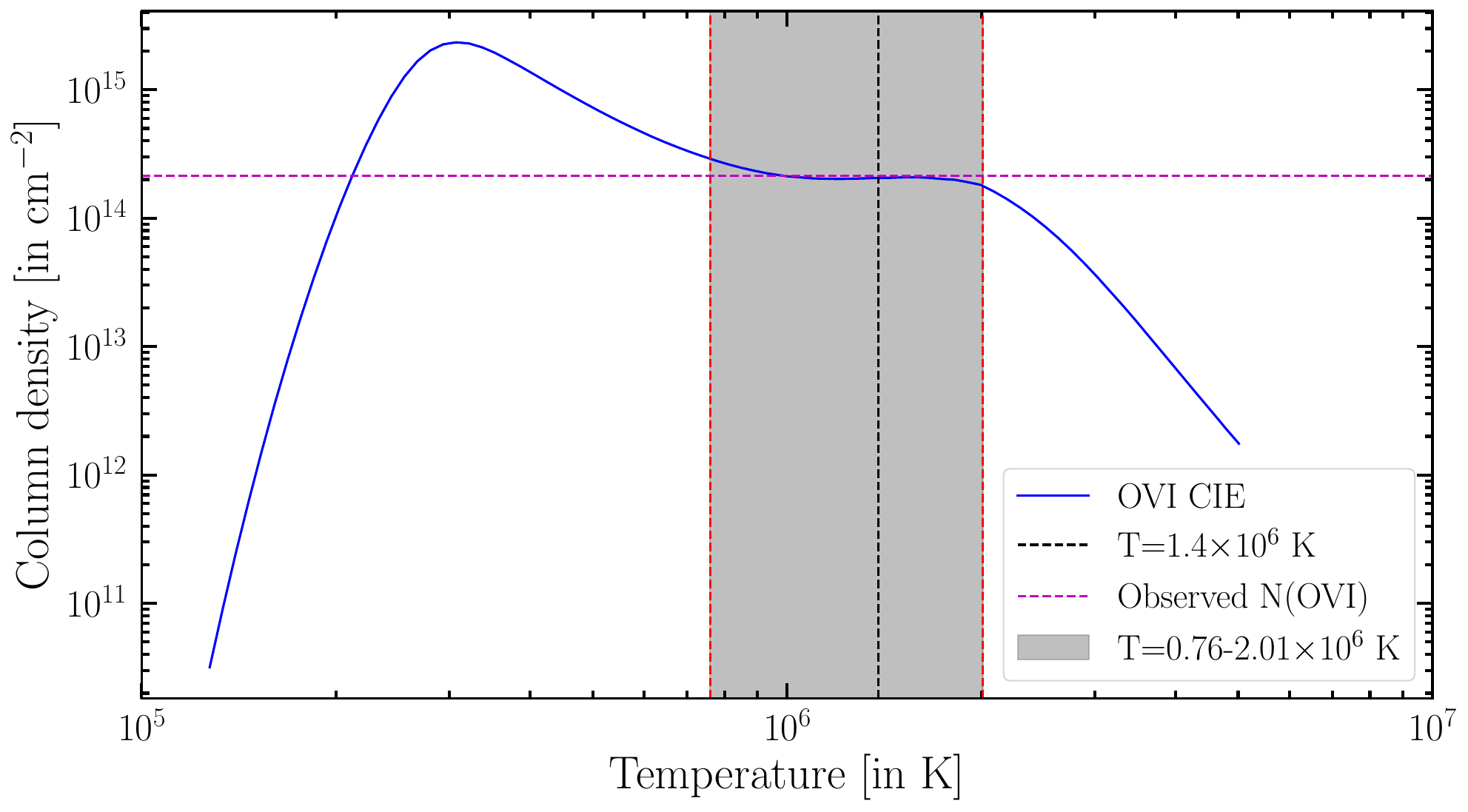}
\caption{Ionization fraction models from \citet{Gnat} for $\log (\textup{Z}/\textup{Z}_{\odot})=-0.50$. The blue curve represents the CIE column density of {\OVI}, respectively, with respect to temperature. The vertical black dashed lines correspond to constant temperature; the gray region shows the 1$\sigma$ temperature range; and the horizontal dashed line corresponds to the observed {\OVI} column density. This shows the calculated temperature is in agreement with the CIE models at the observed {\OVI} column density.}
\label{fig:cie}
\end{center}    
\end{figure}

An alternative is for the {\OVI} to arise in the same gas phase as the BLA. The $b(\textup{H}) = 156^{+24}_{-38}$~{\kms} and $b(\textup{O}) = 42~\pm~5$~{\kms} solves for the temperature of such a phase to be in the range of $T = (0.76 - 2.01) \times 10^6$~K, with a mean of $T = (1.38~\pm~0.62) \times 10^6$~K. The range is obtained from the $1\sigma$ limiting values for the $b$ parameters of the BLA and {\OVI}. In Figure \ref{fig:cie}, we show the collisional ionization equilibrium (CIE) model predictions for {\OVI} based on the models of \citet{Gnat}. Within the temperature range predicted by the different $b$-values of the BLA and {\OVI}, the predictions of the CIE models are more or less constant. At $T \approx 10^6$~K, the ionization fractions of {\CIII} and lower ions are insignificant. The ionization fractions of $f(\OVI) = 1.94 \times 10^{-3}$ and $f(\HI) = 1.49 \times 10^{-7}$ yields an oxygen abundance of $[\textup{O}/\textup{H}]\sim -0.50$ and a total ($\HI~+~\HII$) hydrogen column density of ${\textup{N(H)} = 6.85 \times 10^{20}\:{\cmsq}}$, {which is almost an order of magnitude more than the column density of baryons} in the cooler component of the absorber responsible for the Lyman break. The model results are given in Table \ref{tab:phases}.

As explained earlier, the {\OVI} coinciding in velocity with the Lyman break component at $v \approx 0$~{\kms} is unexplained by the photoionized phase responsible for the low ions. It could be that this {\OVI} is an extension of the hot phase of the gas traced by the BLA and {\OVI}. The corresponding {\Lya} absorption from the trace amounts of {\HI} in this highly ionized gas is masked by the strong absorption from the cooler {\HI} responsible for the Lyman break. The ionization modeling results for the three components in {\HI} are summarized in Table \ref{tab:phases}. In order to determine the origin of the absorbing gas with these properties, we need to observe the quasar field to see if there are some galaxies that are coincident with the absorber.

\section{Galaxies coincident with the absorber}\label{sec:galaxies}

\begin{figure*}
\begin{center}
\begin{minipage}{0.99\textwidth}
\includegraphics[width=\linewidth]{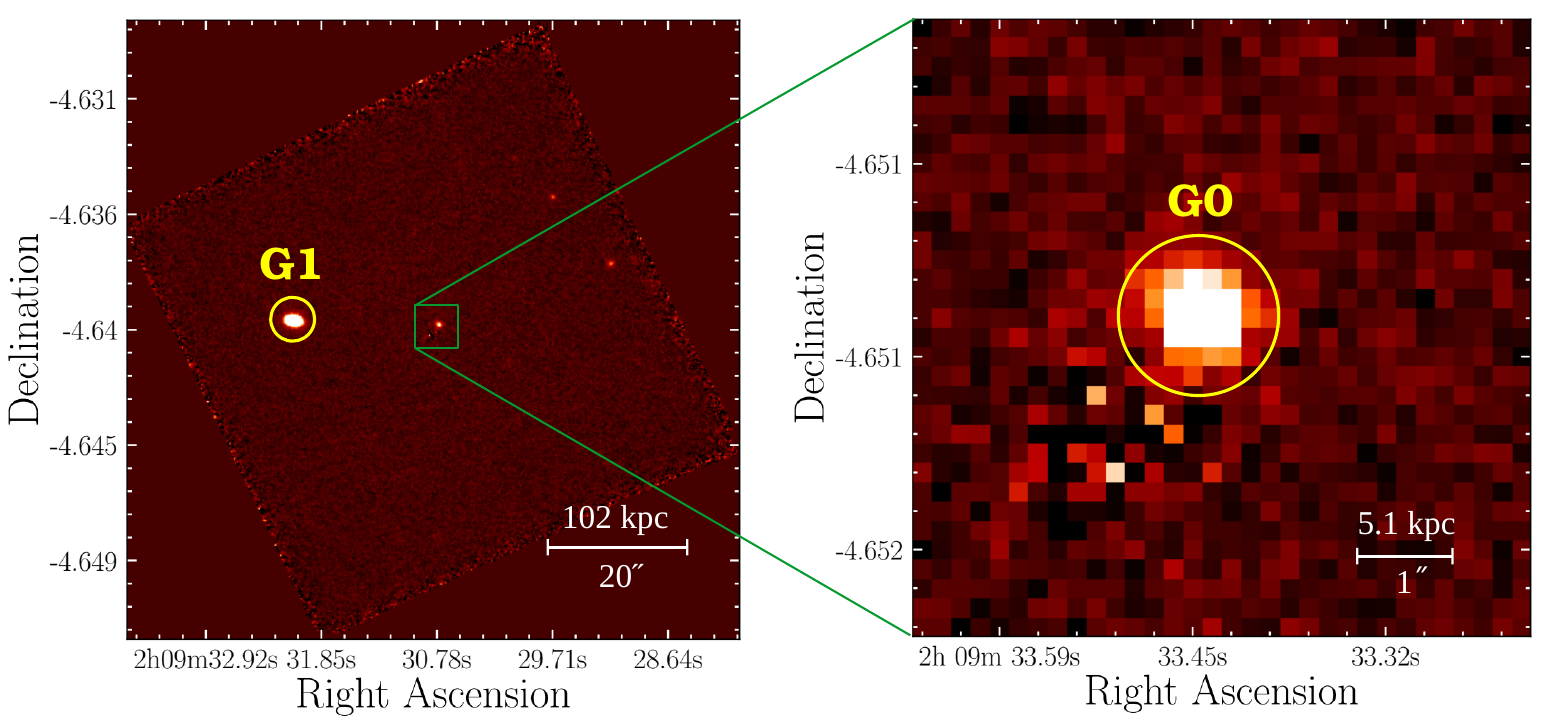}
\end{minipage}
\begin{minipage}{0.49\textwidth}
    \centering
    \includegraphics[width=\linewidth]{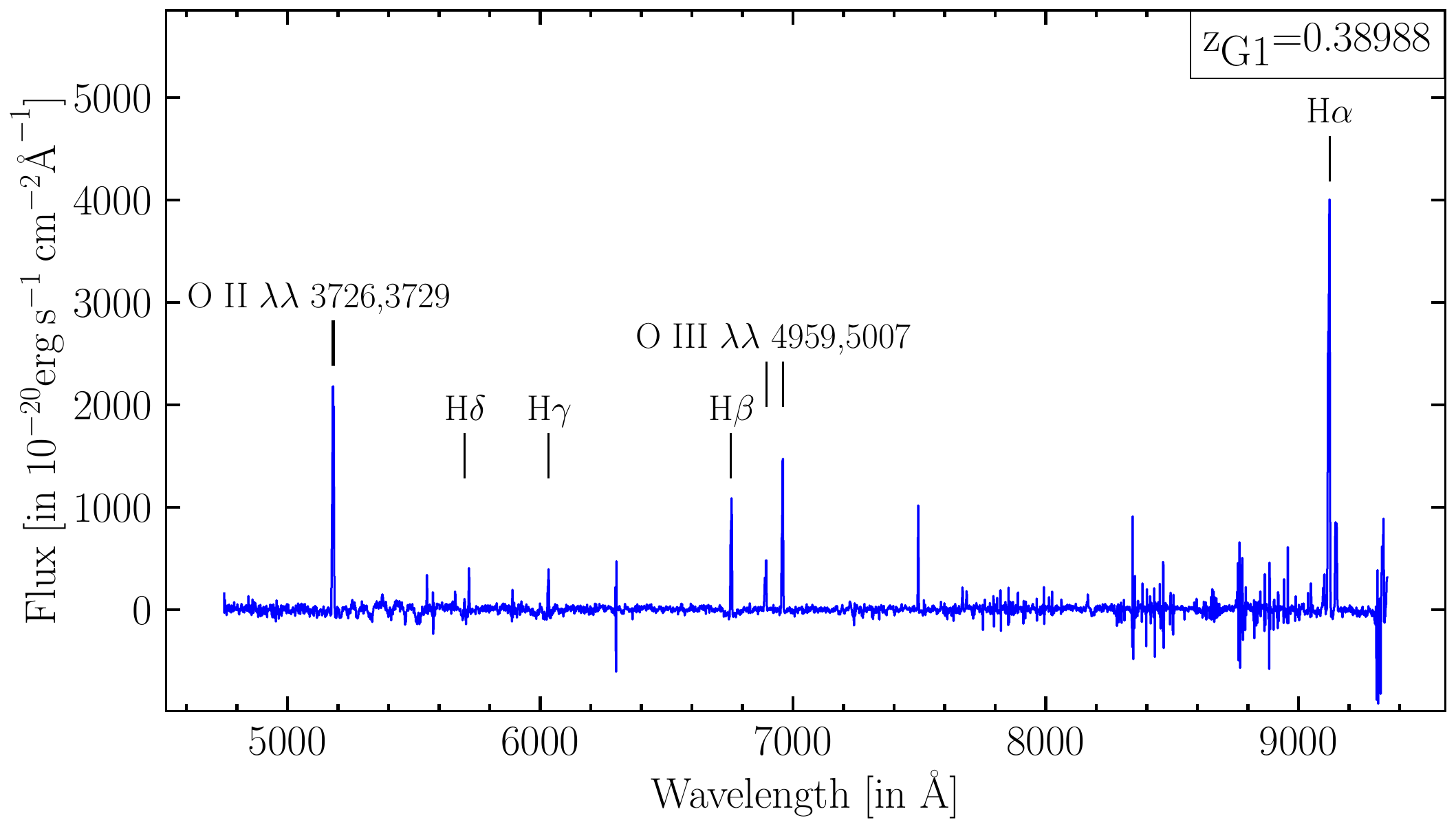}
\end{minipage}
 \begin{minipage}{0.49\textwidth}
    \centering
    \includegraphics[width=\linewidth]{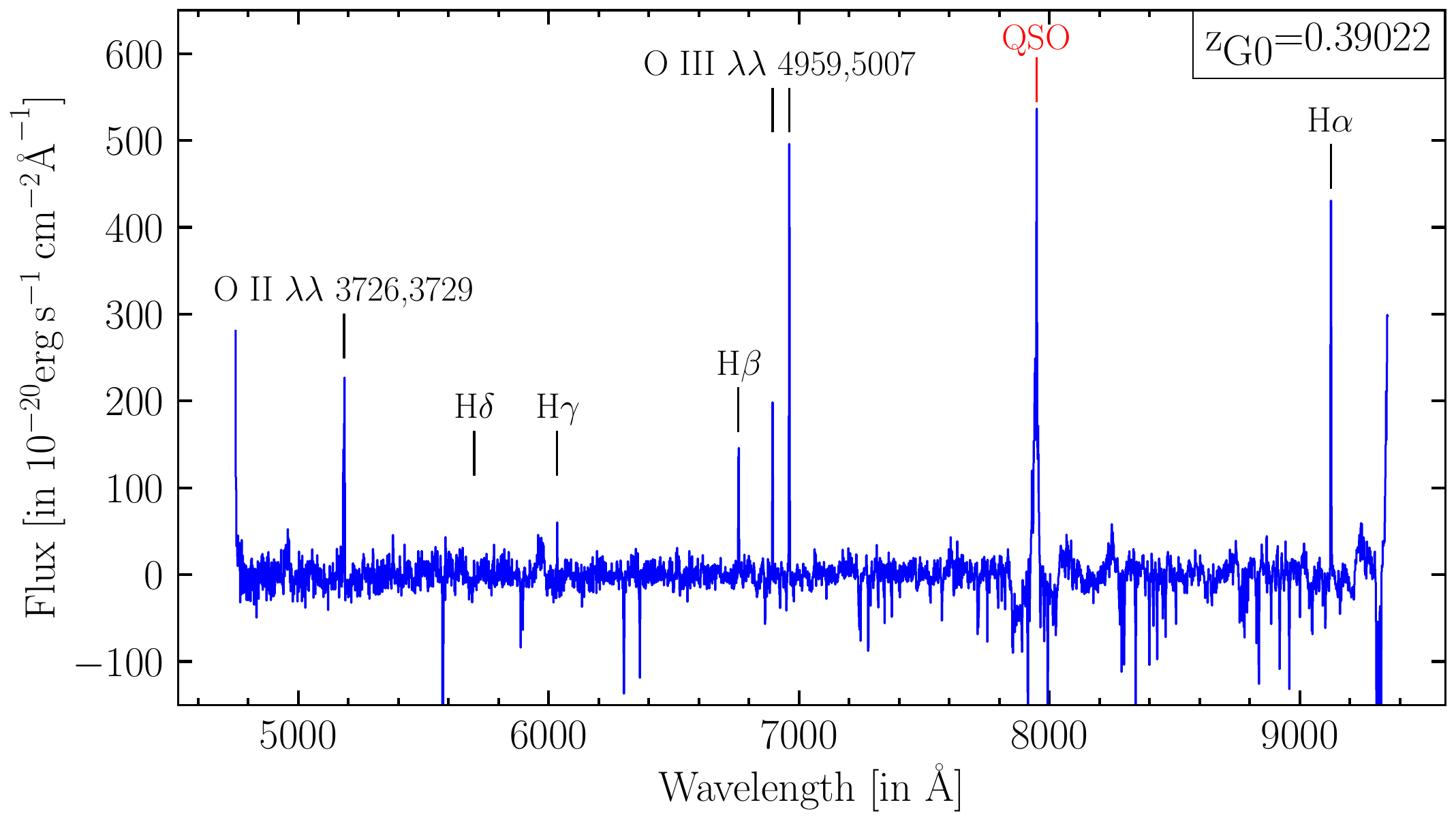}
\end{minipage}
\caption{The top left panel shows a narrow H$\alpha$ emission image at the location of the absorber, obtained from the continuum and quasar PSF subtracted IFU datacube. The top right panel shows the zoomed-in image of galaxy G0, a GOTOQ. The bottom panels show the emission line-rich spectra of G1 and G0 extracted by adding the spaxels covered by them, which were used to find their redshifts that resulted in a value almost the same as the absorber redshift $\sim$0.39. The emission feature labeled ``QSO'' in the G0 spectrum is residual {\OII} emission, after PSF subtraction, from the quasar itself.}
\label{fig:G_spectra}
\end{center}
\end{figure*}

The quasar field is covered by the MUSEQuBES, which is a blind survey for Ly-alpha emitters in the $1^{\prime}\times 1^{\prime}$ field around bright quasars. The MUSE data reduction is described in \citet{Muzahid20}. The survey has identified two galaxies in the foreground field of the quasar with systemic velocities coincident with the absorber. In Figure \ref{fig:G_spectra}, we show the H$\alpha$ narrowband image of the field with the two galaxies labeled G0 and G1. The galaxies are at projected separations of $\rho \approx 12,$ and $104$~kpc, and line of sight velocity separations of $\Delta v \approx -54$~{\kms} and $\Delta v \approx -127$~{\kms}, respectively from the absorber. G0 is in a GOTOQ configuration. The quasar light had to be removed to reveal this galaxy, as explained in Section \ref{sec:ifu}.

\begin{table}
\begin{center}
\caption{Galaxies G0 and G1 parameters from the MUSEQuBES catalog}
\begin{tabular}{c c c} \hline
\textbf{Parameter} & \textbf{G0}  & \textbf{G1}  \\ \hline \hline
\textbf{$\boldsymbol{\textup{z}}$}   & $0.39022$ & $0.38988$ \\
\textbf{$\boldsymbol{\textup{M}_\ast}$ ($\boldsymbol{\textup{M}_\odot}$)}              & $\approx 10^{6}$ & $\approx 10^{10}$ \\
\textbf{SFR ($\boldsymbol{\textup{M}_\odot \: \textup{yr}^{-1}}$)}        & $0.01$  & $1.78$  \\
\textbf{$\boldsymbol{\textup{R}_\textup{vir}}$  (kpc)}             & $31.53$ & $138.93$  \\ 
\textbf{[O/H]}             & $-0.058$ & $0.218$  \\ \hline
\end{tabular}
\label{tab:MUSEQuBES}
\end{center}
\end{table}

\begin{table*}
\renewcommand{\arraystretch}{1.5}
\centering
\caption{The results from  GalPaK$^{3\textup{D}}$ for G1}
\begin{tabular}{l c c c c} \hline
\textbf{Parameter}      & \textbf{[$\boldsymbol{\OII}$]}  & \textbf{H$\boldsymbol{\beta}$}  & \textbf{[{\OIII}] 5007}  & \textbf{H$\boldsymbol{\alpha}$}                  \\
\hline \hline
\textbf{$\boldsymbol{\textup{Flux}}$ ($\boldsymbol{10^{-16}\: \textup{erg}\: \textup{s}^{-1}\: \textup{cm}^{-2}}$)}          & $ 1.76~\pm~0.008$  & $ 0.64~\pm~0.003$  & $ 0.84~\pm~0.003$  & $ 2.94~\pm~0.004$    \\
\textbf{$\boldsymbol{\textup{R}_{1/2}}$ (kpc)} & $ 4.45~\pm~0.03$  & $ 3.86~\pm~0.03$  & $ 4.13~\pm~0.03$  & $ 3.37~\pm~0.02$    \\
\textbf{Inclination} & $71.3^{\circ}~\pm~0.3$  & $76.7^{\circ}~\pm~0.3$  & $78.6^{\circ}~\pm~0.2$  & $78.9^{\circ}~\pm~0.1$ \\ 
\textbf{P.A.} & $262.4^{\circ}~\pm~0.2$  & $263.1^{\circ}~\pm~0.2$  & $262.9^{\circ}~\pm~0.2$  & $263.3^{\circ}~\pm~0.1$ \\ 
\textbf{$\boldsymbol{\alpha}$ } & $1.1^{\circ}~\pm~0.2$ & $1.8^{\circ}~\pm~0.2$ & $1.6^{\circ}~\pm~0.2$ & $2.0^{\circ}~\pm~0.2$ \\ 
\textbf{$\boldsymbol{\textup{V}_\textup{max}}$ ($\boldsymbol{\textup{km}\: \textup{s}^{-1}}$)} & $128~\pm~1$  & $127~\pm~1$  & $124~\pm~1$  & $130~\pm~1$ \\ 
\textbf{$\boldsymbol{\textup{R}_\textup{vir}}$ (kpc)} & $ 152 $ & $ 150 $ & $ 146 $ & $ 153 $    \\
\textbf{$\boldsymbol{\textup{M}_\textup{dyn} (<\boldsymbol{\textup{R}_{1/2}})}$ ($\boldsymbol{10^{10}\:\textup{M}_\odot}$)} & $1.71$ & $1.45$ & $1.51$ & $1.32$ \\ 
\textbf{$\boldsymbol{\textup{M}_\textup{h}}$ ($\boldsymbol{10^{11}\:\textup{M}_\odot}$)} & $6.83$ & $6.59$ & $6.14$ & $7.03$ \\ 
\textbf{$\boldsymbol{\textup{M}_\ast}$ ($\boldsymbol{10^{10}\:\textup{M}_\odot}$)} & $2.91$ & $2.79$ & $2.55$ & $3.01$ \\ 
\hline
\end{tabular}
\flushleft{\tablecomments{{This table shows the parameters obtained from GalPaK$^{3\textup{D}}$ for different emission-line datacubes of the galaxy G1. (i) The flux in each emission band; (ii) Half-light radius; (iii) the inclination of the galaxy relative to the sky plane; (iv) Position angle; (v) Azimuthal angle; (vi) Maximum rotational velocity; (vii) Virial radius; (viii) Dynamical mass within the half-light radius; (ix) Halo mass of the galaxy; (x) Stellar mass of the galaxy. It is observed that the values of the parameters remain consistent for different emission bands.}}}
\label{tab:G1_pars}
\end{table*}

\begin{figure*}
\begin{center}
\begin{minipage}{0.49\textwidth}
    \centering
    \includegraphics[width=0.77\linewidth]{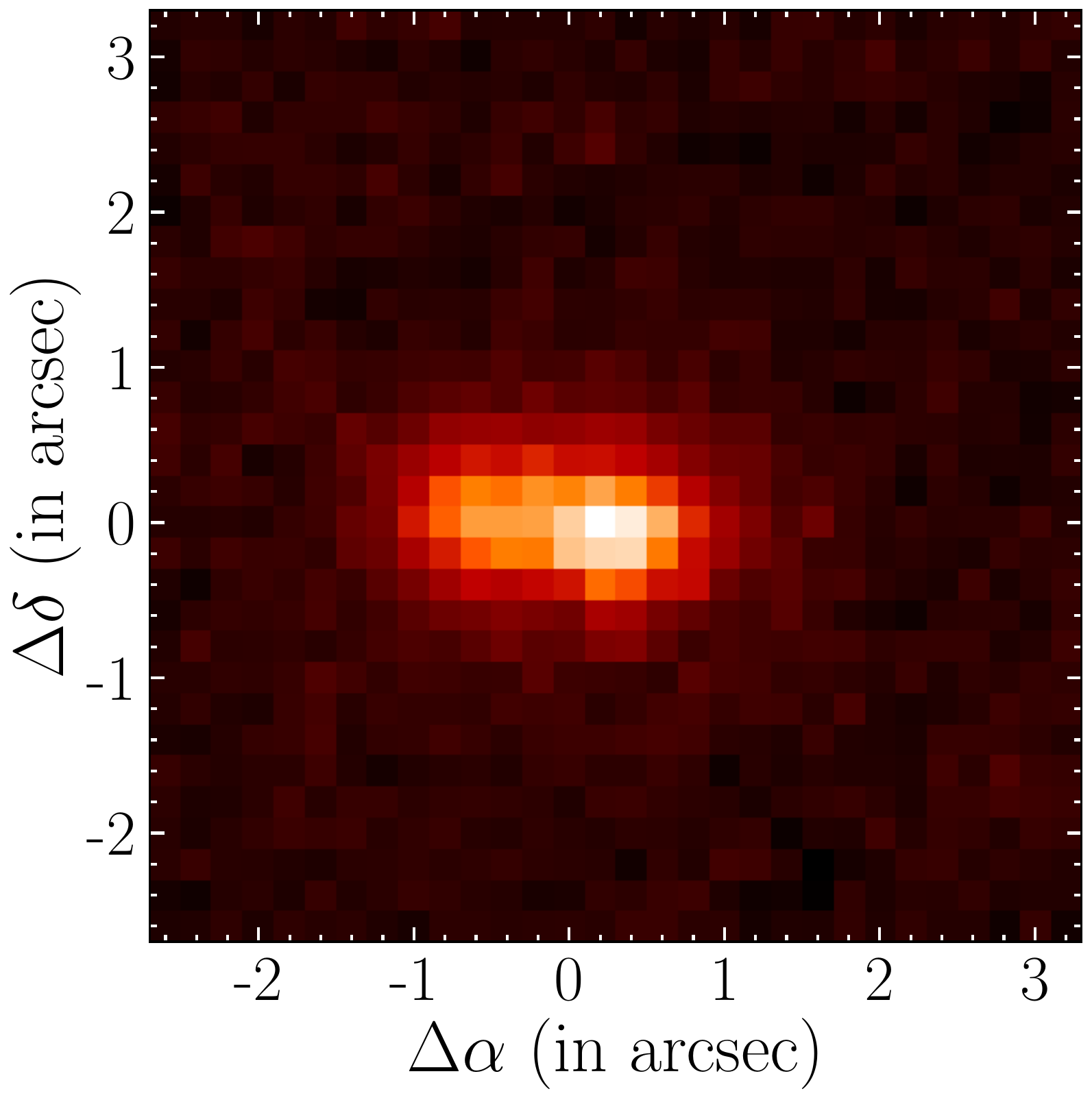}
\end{minipage}
 \begin{minipage}{0.49\textwidth}
    \centering
    \includegraphics[width=\linewidth]{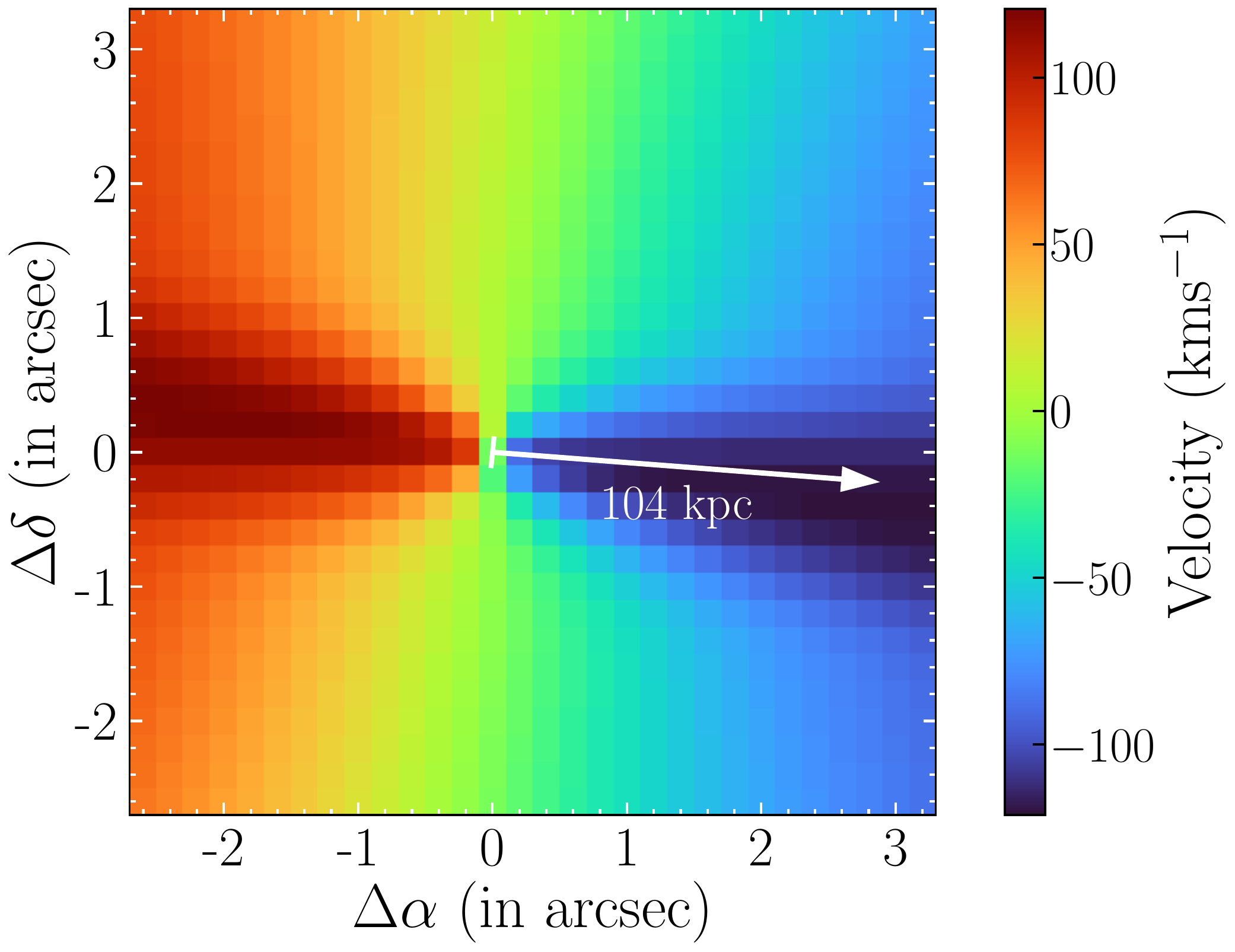}
\end{minipage}
\caption{The left panel shows a flux map of the galaxy G1 from the {\OII} narrow band datacube of spatial dimensions $6^{\prime\prime}\times 6^{\prime\prime}$. The right panel shows the velocity map of this galaxy obtained from the GalPaK$^{3\textup{D}}$ analysis. The disk-type structure of the galaxy is clear from this map. The white arrow points towards the location of the quasar sight line at a distance of 104 kpc from the center of the galaxy.}
\label{fig:G1_analysis}
\end{center}    
\end{figure*}

\begin{figure*}
\begin{center}
\includegraphics[width=0.6\linewidth]{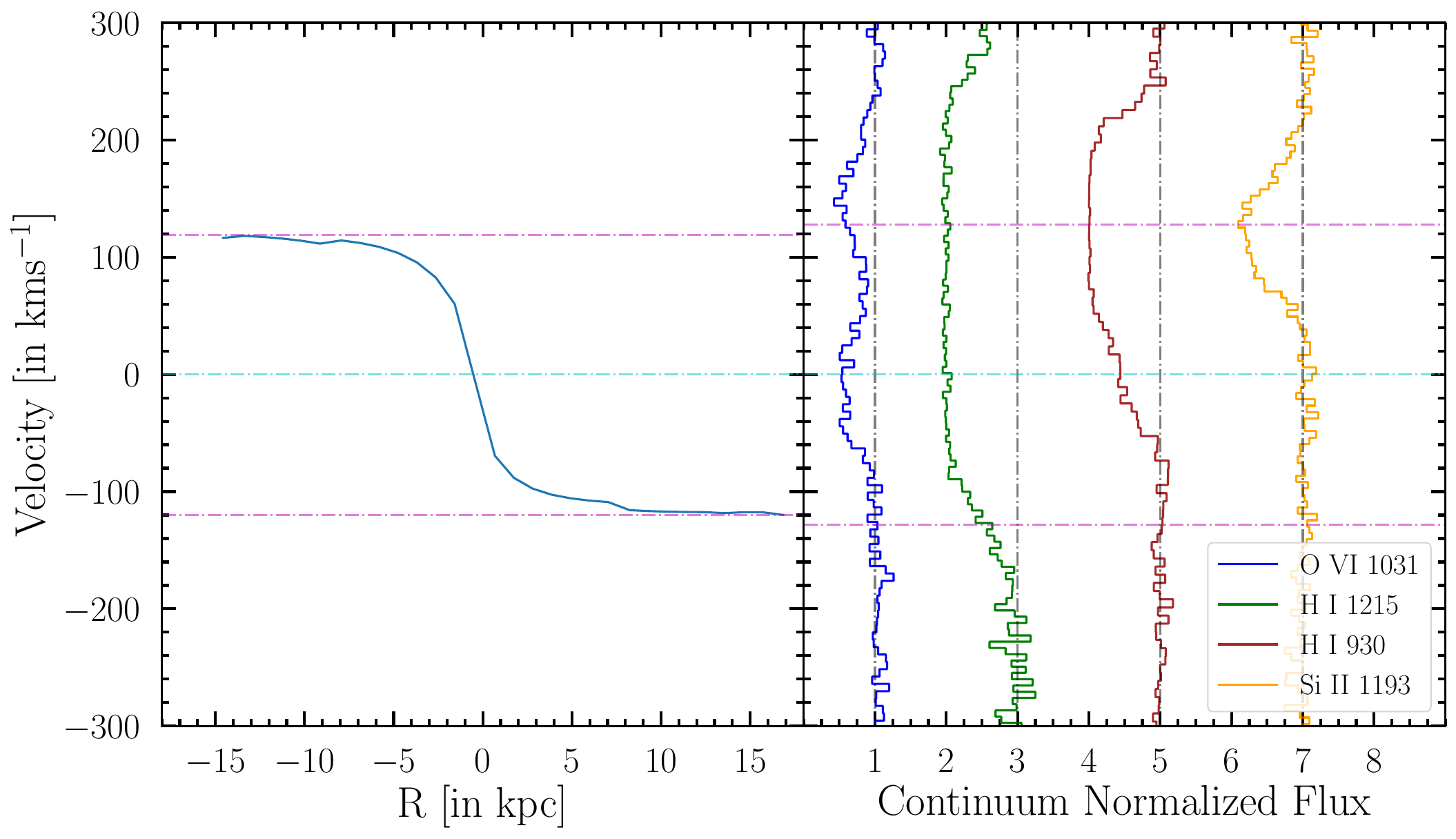}
\caption{The rotation curve of galaxy G1 is shown on the \textit{left} panel and the velocity profiles of {\HI}, {\SiII}, and {\OVI} lines in  the \textit{right} panel. The $v = 0$~{\kms} in the vertical axis corresponds to $z=0.38988$, the redshift of G1. The horizontal axis in the \textit{left} panel is projected galactocentric distance, and in the \textit{right} panel, it is the continuum normalized flux, offset deliberately to view the absorption features created by different species clearly. The low ionization gas, traced by {\SiII} and the LLS component of {\HI}, is offset from the systemic velocity of the galaxy towards the positive velocity side (see Figure \ref{fig:G1_analysis}). The line-of-sight velocity of the absorbing material (QSO line of sight, which is positioned at R$\sim+104$ kpc) is thus opposite to the rotational direction of G1's disk.}
\label{fig:gal_kin}
\end{center}
\end{figure*}

Galaxy G0 appears to be a dwarf galaxy extending $\sim 1.6^{\prime\prime}\times 1.2^{\prime\prime}$ in the {\OII} narrowband image, corresponding to a physical size of $8.5$~kpc~$\times~6.4$~kpc. The $M_* \approx 10^6$~M$_\odot$ stellar mass given by the MUSEQuBES catalog (see Table \ref{tab:MUSEQuBES}) is consistent with this. In comparison, galaxy G1 is a massive star-forming galaxy with an angular size that is approximately three times larger than G0. The absorber location, given by the line of sight to the background quasar, and galaxy G0 are both within G1's virial halo ($\rho/R_\textup{vir} \approx 0.7)$. Thus, G0 could be a dwarf satellite to G1. 

We use the GalPaK$^{3\textup{D}}$ algorithm \citep{Bouche15A,Bouche15} to derive the morphological and kinematic properties of galaxy G1. A similar analysis could not be performed on G0, as the galaxy is barely resolved with a physical size that is only a factor of two larger than the seeing FWHM of $3.5$~kpc. In addition, the possibility of contamination of the spaxels corresponding to G0 from the residual of quasar PSF subtraction cannot be ruled out. The morpho-kinematic analysis is, therefore, restricted to G1. 

For G1, we use four continuum-subtracted sub-cubes of spatial dimensions $6^{\prime\prime}\times 6^{\prime\prime}$, centered on the galaxy and cropped around the emission features [{\OII}]~$\lambda \lambda$ 3726, 3729~{\AA}, $\H\beta$, [{\OIII}] 5007~{\AA}, and $\H\alpha$. GalPaK$^{3\textup{D}}$ explores the morpho-kinematic parameter space using a disk parametric model with ten free parameters and the MCMC algorithm. Table \ref{tab:G1_pars} lists the properties of galaxy G1 given by the GalPak$^{3\textup{D}}$ algorithm. Galaxy G1's flux and velocity maps are shown in Figure \ref{fig:G1_analysis}. The velocity map and the inclination angle clearly indicate that G1 is a disk galaxy, oriented nearly edge-on. The virial radius, the dynamical mass within the half-light radius, and the halo mass listed in Table \ref{tab:G1_pars} were calculated from the inclination-corrected maximum rotation velocity using the expressions given in \citet{Bouche}. The stellar mass of the galaxy was estimated from the stellar-to-halo mass scaling relation of \citet{Girelli}. 

{The metallicities for the galaxies by the MUSEQuBES survey were determined using the R23 method described in \citet{Kobulnicky99}. We have used different indicators to calculate the metallicity of the galaxies, e.g., R23, N2, R3, O32, R2, O3N2 \citep{Maiolino08}. The metallicity for G1 comes out to be slightly super-solar in all these cases (log (Z/$\textup{Z}_{\odot}$) $\approx +0.1$), which is in agreement with the MUSEQuBES catalog value. But for G0, the metallicity cannot be estimated robustly because of the contamination of the emission flux from the galaxy spectrum due to the residual light from quasar left after the PSF subtraction. However, we find that the metallicity to be sub-solar in all the cases. The extremely weak {\NII} 6583 line in its spectrum also suggests that G0 is a metal-poor galaxy \citep{Kobulnicky99}.} {These ISM abundances are higher than the metallicity estimated from ionization modeling for the cooler photoionized gas traced by the LLS. The Lyman limit absorption thus might not be directly tracing recent winds or interstellar gas stripping from either galaxy, even though the absorber resides within the circumgalactic envelope of both galaxies.} CGM metallicities are typically lower than galaxy metallicities, as seen in larger samples of absorber-galaxy studies, with mean offsets between metallicities of $1/10$th solar and more (e.g., \citet{Kacprzak19}). The metallicity differences point to the diverse origins of clouds in the halos of galaxies and the complex mixing of gas expelled or accreted as it moves through the multiphase gas reservoirs of the CGM. {The following section discusses the possible processes which led to the formation of this absorbing cloud.} 

\section{Physical origin of the absorber}\label{sec:origin}

The line of sight to the quasar is traversing the halo environment of both galaxies. The absorption, therefore, cannot be unambiguously assigned to the circumgalactic medium (CGM) of one. For optically thick {\HI} absorbers, it is common to find two or more galaxies of different masses coincident with the absorber \citep{Peroux19,Hamanowicz20,Kulkarni22,Weng23}, with multiple galaxies contributing to the different gas phases of the absorber \citep{Nateghi21}. This makes host galaxy identification difficult. Galaxies G0 and G1 are close enough in projected separations ($\rho/R_\textup{vir} \approx 0.4$ and $0.7$) to contribute to the absorption. 


Simulation studies suggest that massive halos ($10^{10} \lesssim M_h \lesssim 10^{12}~$M$_\odot$) have nearly three orders of magnitude larger covering-fraction of high column density {\HI} gas within their projected virial radii compared to the covering fraction of similar gas around dwarf systems \citep{Hafen}. This indirectly implies that when multiple galaxies are present at the absorber location, there is a higher probability of the absorbing gas being gravitationally bound to the central massive galaxy than any of its lower-mass companions \citep{Kohler07}. A trend suggesting this is seen in observations in the form of a positive correlation between the stellar mass of the nearest galaxy and the line of sight's impact parameter to it, as massive galaxies invariably possess a much more extended and denser envelope of gas than dwarfs \citep{Kulkarni22}. Meanwhile, \citet{Weng22} contend that the scatter in the well-established correlation between absorption line strength and impact parameter to the nearest galaxy might be the result of the absorber being associated with more than one galaxy. The correlation invariably attributes the absorption to the nearest galaxy, which may not always be the dominant or the only contributor to the absorption. Where multiple galaxies are present, dynamical interactions can alter the physical extent of the CGM through the redistribution of gas within the overlapping haloes. This enhances the covering fraction of hydrogen and metals in the overall environment surrounding those galaxies \citep{Hani18}. Therefore, rather than isolating the LLSs to G0 or G1, we may have to consider the combined environment of both the galaxies where the absorption is originating. 

The inclination and the azimuthal angle (see Table \ref{tab:G1_pars}) show galaxy G1 to be nearly edge-on and the quasar line of sight within $R_\textup{vir}$ probing a region along its extended major axis. The GOTOQ galaxy G0 is also within $R_\textup{vir}$ of G1. The sSFR of $\approx 0.18$~Gyr$^{-1}$ for G1 suggests the likelihood of G1 experiencing outflows driven by stellar feedback. However, the azimuthal angle, or the lower metallicity of the absorbing gas compared to the galaxy, does not favor the line of sight directly probing outflows. A higher cross-section for absorption beyond the stellar disk is possible if the disk has a warped structure formed either from the accretion of recycled wind material from past outflows or tidal interactions with nearby satellites \citep{Garcia02,Rahmani18,Semczuk20}. The existence of the dwarf companion G0 supports the latter scenario. The metallicity of $\log (Z/Z_\odot) \approx -0.98$ that we obtain for the cooler phase of the gas associated with the Lyman limit is consistent with the median metallicity for LLSs from recent surveys \citep{Zahedy21}. The metallicity aligns with some of the recent cosmological simulations where LLSs with $\log (Z/Z_\odot) \gtrsim -1$~dex effectively trace recycling winds, ancient outflows, or tidally stripped material as opposed to pristine gas accreting onto galaxies (e.gs., \citet{Oppenheimer10,Ford14}). The mixing of such displaced gas with the metal-poor regions of the CGM can lower the overall metallicities, especially if the mixing is inhomogeneous, confining metals to patchy zones \citep{Schaye07,Sankar20}. 

The warm/hot phase of the gas traced by the {\OVI}-BLA is similar in properties to several multiphase CGM absorbers reported earlier \citep{Savage10,Savage11,Narayanan18}. {\OVI} is known to be widespread around star-forming galaxies \citep{Tumlinson11,Werk_16,Muzahid14,Muzahid,Tchernyshyov22}, where it typically traces spatially extended warm/hot plasma \citep{Zahedy21}. If associated with the coronal regions of galaxy G1, the baryonic mass ({\HI} + {\HII}) entrenched in this warm/hot phase can be as high as $M_\textup{b} = \pi r^2 \mu m_{\H} N(\H) \approx 10^{11}$~M$_\odot$\footnote{In this estimate based on \citet{Fox13}, we have assumed a {\OVI} covering fraction of unity out to $r = 104$~kpc around G1, in line with the covering fraction estimates of \citet{Tumlinson11} for star-forming galaxies. The $r = 104$~kpc is the impact parameter of G1 with the quasar line of sight. We further assume a mean mass of $\mu = 1.3$ in units of $m_{\H}$, and an estimated total {\H} column density of $6.85 \times 10^{20}$~{\cmsq} obtained from the CIE modeling.}. This is comparable to the mass contained in the $T \approx 10^4$~K photoionized Lyman limit phase traced by the lower ionization metals and the bulk of the {\HI}. Such reservoirs of baryons in the warm/hot phase are common to LLSs tracing the CGM \citep{Fox13}. The above estimation of baryonic mass is an upper limit if the {\OVI} is tracing filamentary structures such as in gas accretion or conical structures such as in outflows rather than the assumed spherical geometry \citep{Kacprzak19}, though {\OVI} is known to be widespread with a nearly $100$\% covering fraction within $R_\textup{vir}$ of $\gtrsim L^*$ galaxies (e.g., \citet{Lehner20}). 

The origin of circumgalactic {\OVI} is perhaps best understood in the Galactic halo and around galaxies in the Local Group (e.g., \citet{Friedman00,Wakker03,Lehner07,Lehner20}). In HVCs, the {\OVI} is collisionally ionized at the sheet-like transition temperature layers that form between the cooler ($T \approx 10^4$~K) photoionized gas and the fully ionized hot ($T \gtrsim 10^6$~K) coronal halo that forms an exterior medium (e.g., \citet{Fox04,Fox05}). The cooler phase, where the hydrogen is predominantly neutral, can have column densities corresponding to LLSs $\textup{N}(\HI) \gtrsim 10^{18} \: \textup{cm}^{-2} $  where they are also identified in their 21-cm {\HI} emission \citep{Fox06,Wakker12}. The $z \approx 0.39$ absorber described in this paper could be an analog of such high-velocity gas formed from the dynamical interaction between the galaxy G1 and its dwarf companion G0. A line of sight piercing a high-velocity cloud complex will pass through the transition temperature outer layers at least twice. The two distinct components of the {\OVI} absorption separated by $\approx 160$~{\kms} in the absorber frame can be such a layer of collisionally ionized plasma at the interface between the gas responsible for the LLSs, and the more extended coronal envelope of hot gas around G1 (the more massive galaxy among the two). The kinematic offset between the low ionization gas and the {\OVI} phase and the radial velocity difference of $|\Delta v| \approx 127$~{\kms} between the absorber and G1 favors this possibility.     

Finally, we compare the line-of-sight kinematics of this circumgalactic gas with the kinematics of the G1 galaxy. Figure \ref{fig:gal_kin} shows G1's rotation curve obtained from GalPaK$^{\mathrm{3D}}$ analysis and the profiles of {\HI}, {\OVI}, and {\SiII} lines centered on the systemic redshift of the galaxy. The bulk of the absorption is traced by the low ionization species (represented by {\SiII}), and the LLS lies to one side of the galaxy's velocity, as is expected if the absorption is tracing material that is farther out in the disk. However, the absorber does not share the velocity sign of the galaxy's rotation (see Figure \ref{fig:G1_analysis}), implying that the absorbing gas {may not be} co-rotating with the disk. Circumgalactic gas in its final stages of accretion onto galaxies is expected to form an extended cold flow disk with velocities in the same direction as the galaxy's rotation \citep{Stewart11,vandevoort11,Ho17,Zabl19,Weng23}. Though aligned with the extended major axis of G1, the  {line-of-sight} kinematics of the absorber relative to the galaxy does not suggest such co-rotating accretion. At the same time, the possibility of the absorber being counter-rotating gas accreting onto rotating disks cannot be excluded (e.g., \citet{Dyda15}).  In any case, the ionization structure of the absorber and its line-of-sight velocities relative to the galaxies are consistent with high-velocity gas moving through the circumgalactic medium of either galaxy.

\section{Summary of Results}\label{sec:sum}

In this work, we analyze a Lyman limit absorber complex at $\rm z=0.39047$ detected in the far-UV $HST$/COS spectrum of the background QSO, FBQS~J$0209-0438$. The quasar field is also observed by the MUSE IFU instrument on $VLT$. The following are the major conclusions from the analysis:

\begin{enumerate}
    \item The absorption system at $z = 0.39047$ is detected in {\HI} (full range of Lyman series lines), {\CII}, {\NII}, {\NIII}, {\SiII}, {\CIII}, and {\OVI}. The Lyman break associated with this system has a ${\tau}_{912\:\textup{\AA}} > 7.1$, and a corresponding $\log [\textup{N}(\HI)/\textup{cm}^{-2}] > 18$.     
    \item Simultaneous Voigt profile fitting of the Lyman lines results in two components. One of the components has a $\log [\textup{N}(\HI)/\textup{cm}^{-2}] \approx 18.6$ at $v \approx -2$~{\kms} in the absorber frame, causing a Lyman limit break.  The second component has a column density that is three orders of magnitude lower, constrained by the unsaturated absorption in the higher-order Lyman lines. 
    \item The two-component fit to the Lyman series lines leaves a residual in the blue end of the {\Lya} profile, which can be explained by introducing a third shallow component. The third component is a BLA with $\log [\textup{N}(\HI)/\textup{cm}^{-2}] \approx 14.0\pm~0.2$, $b(\H) = 156~\pm~30$~{\kms}, coinciding in velocity with the nearest {\OVI}. 
    \item The low ions are detected only at the velocity of the strong {\HI} component responsible for the Lyman limit break, whereas the {\OVI} has a different kinematic profile with two distinct components separated in velocity by $\Delta v \approx 160$~{\kms} from each other, with the blueward component within $1\sigma$ of the BLA velocity centroid.  
    \item Photoionization equilibrium modeling with a Bayesian MCMC approach shows the component corresponding to the Lyman limit break to be consistent with gas at {$T \approx 1.37 \times 10^4$ K, $n_{\H} \approx 6.46 \times 10^{-3} \: \cc$, and log$(\textup{Z}/\textup{Z}_{\odot}) = -0.98 \pm 0.07$}. This phase under-predicts the {\OVI} that is nearest in velocity by six orders of magnitude but explains the observed column densities of the low ions. 
    \item The different $b$-values of the BLA and the nearest {\OVI} solves for a temperature of $T = (0.76 - 2.01)~\times~10^6$~K. Within this range, collisional ionization equilibrium models recover the observed {\OVI} at [O/H] $\approx -0.5$, with a total hydrogen column density that is about an order of magnitude more than the cooler phase of the gas responsible for the Lyman limit break. 
    \item The MUSEQuBES survey has identified two galaxies; G0 at projected physical and velocity separations of $\rho \approx 12$~kpc ($\rho/R_\textup{vir} \approx 0.4$), $|\Delta v| \approx 54$~{\kms} from the absorber, overlapping with the quasar PSF, and G1 at $\rho \approx 104$~kpc ($\rho/R_\textup{vir} \approx 0.7$), $|\Delta v| \approx 127$~{\kms} (see Figure \ref{fig:G_spectra}). The projected separations are less than the virial radii of both galaxies. The line of sight could be probing gas within the merged halos of both galaxies.
    \item Galaxies G0 and G1 have dust-extinction corrected SFRs of $0.01$ and $1.78$~M$_\odot$~yr$^{-1}$, respectively. G1 is a luminous star-forming galaxy with a stellar mass of $M_* \approx 10^{10}$~M$_\odot$. G0 is consistent with being a dwarf galaxy with $M_* \approx 10^{6}$~M$_\odot$, and is possibly a satellite of G1. The line of sight to the background quasar passes through a region that is nearly aligned with the projected major axis of G1. 
    \item {The metallicity inferred from ionization modeling of the photoionized Lyman limit component of the absorber is lower than the [O/H] of the galaxies G0 and G1 by factors of $\approx 1.6$ and $\approx 17$, respectively; whereas for the collisionally ionized BLA-{\OVI} phase the metallicity derived from the column densities of {\OVI} and {\HI} is higher than the [O/H] of G0 by a factor of $\approx 2$ and lower than that of G1 by a factor of $\approx 6$.} 
    \item The ionization structure of the absorber, along with its proximity and orientation relative to the galaxies, implies that the absorption likely originates from high-velocity gas within the circumgalactic environment of either galaxy, with the {\OVI} tracing transition temperature layers between the cooler phase of the high-velocity gas giving rise to the LLS, and the hot coronal halo.

\end{enumerate}

\begin{acknowledgments}
Some of the data presented in this paper were obtained from the Mikulski Archive for Space Telescopes (MAST) at the Space Telescope Science Institute. The specific observations analyzed can be accessed via \dataset[https://archive.stsci.edu/missions-and-data/hsla]{https://archive.stsci.edu/missions-and-data/hsla}. STScI is operated by the Association of Universities for Research in Astronomy, Inc., under NASA contract NAS5–26555. Support to MAST for these data is provided by the NASA Office of Space Science via grant NAG5–7584 and by other grants and contracts.
Dr. Vikram Khaire is supported through the INSPIRE Faculty Award (No. DST/INSPIRE/04/2019/001580) of the Department of Science and Technology (DST), India. We would like to thank the MUSEQuBES community for their help.
\end{acknowledgments}



\bibliography{main}{} 
\bibliographystyle{aasjournal}

\label{lastpage}

\begin{figure}
\appendix
\begin{center}
\includegraphics[width=0.82\linewidth]{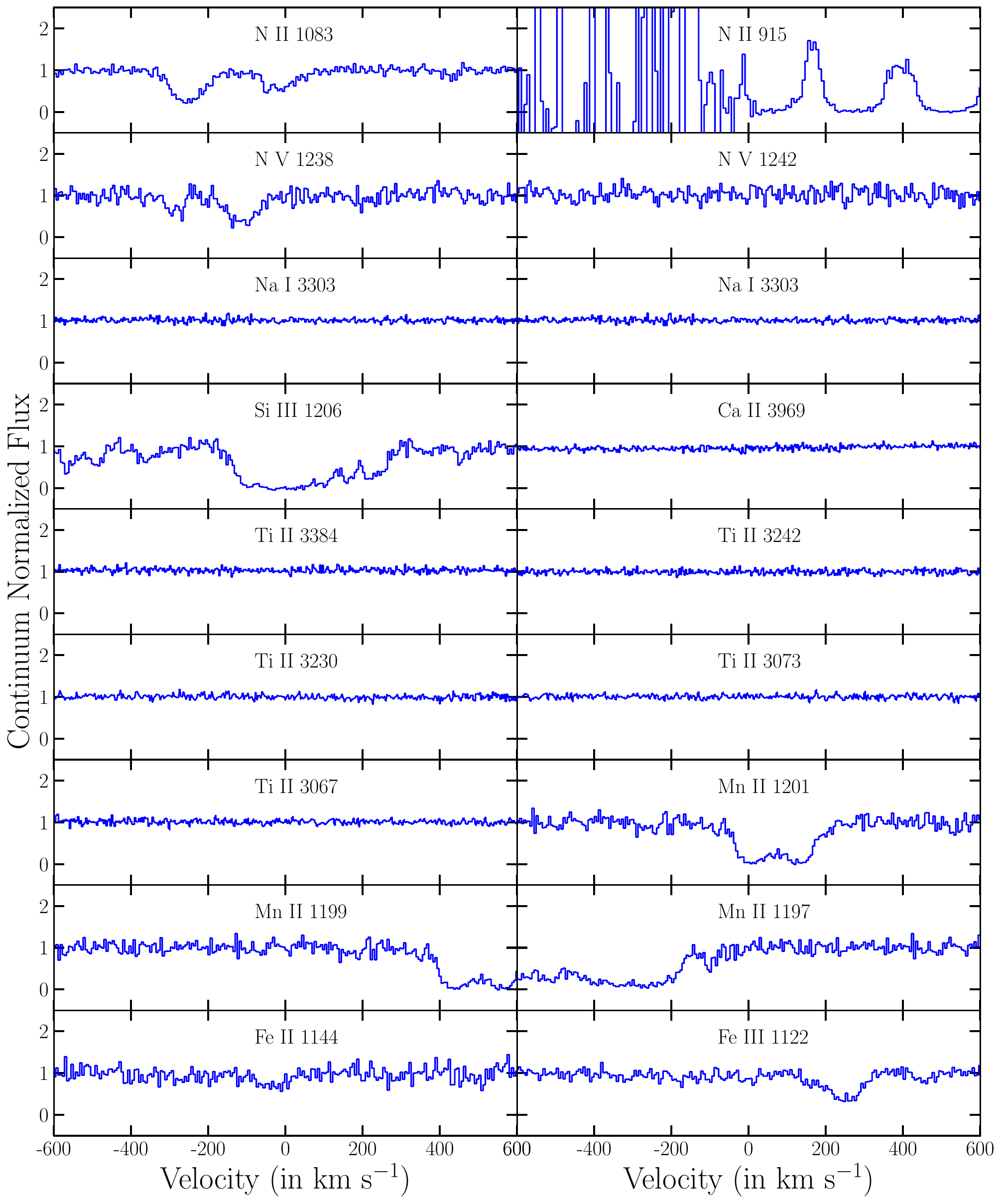}
\caption{{The horizontal axis is the velocity in the rest-frame of the absorber, where $v = 0$~{\kms} corresponds to $z = 0.39047$. This system plot shows the spectra of the quasar obtained from HST/COS and Keck/HIRES, at the expected locations of a few undetected lines. The profile seen at the expected location of the line {\SiIII} 1206 is a result of contamination from {\OIV}~$787$ absorption associated with the quasar and {\HI} 918 from an absorber at z $\sim$ 0.83 \citep{Tejos}}.}
\label{fig:sys2}
\end{center}    
\end{figure}

\end{document}